\documentclass[12pt]{article}
\pdfoutput=1
\usepackage{amsmath,amssymb,amsfonts}
\usepackage{a4wide,epsfig,psfrag,scalefnt}
\usepackage[dvipsnames]{xcolor}
\usepackage{braket}
\usepackage{placeins}
\usepackage{breqn}
\usepackage{slashed}
\usepackage{enumitem}
\usepackage[numbers,sort&compress]{natbib}
\usepackage{caption}
\usepackage{subcaption}

\graphicspath{ {figs/} }

\allowdisplaybreaks

\newcommand{\dflag}[1]{_{\delta^{#1}}}
\newcommand{\pert}{{\mathrm{pert}}}
\newcommand{\kin}{{\mathrm{kin}}}

\newcommand{\OS}{{\mathrm{OS}}}
\newcommand{\GeV}{{\mathrm{GeV}}}

\begin{document}

\begin{flushright}

  P3H-22-048, TTP22-028
\end{flushright}

\begin{center}

  \vspace*{5em}

  {\bf\Large\boldmath A first glance to the kinematic moments \\
  of $B \to X_c \ell \nu$ at third order}\\[15mm]

  \setlength {\baselineskip}{0.2in}
  {\large  Matteo Fael, Kay Sch\"onwald and Matthias Steinhauser}\\[5mm]
  {\it Institut f\"ur Theoretische Teilchenphysik, 
    Karlsruhe Institute of Technology (KIT),\\
  76128 Karlsruhe, Germany}\\[3cm] 

\end{center} 

\centerline{\bf Abstract}
We study the impact of third-order QCD corrections for several kinematic moments
of the inclusive semileptonic $B$ decays, to  first order in the $1/m_b$ expansion.
We consider the first four moments of the charged-lepton energy
$E_\ell$ spectrum, the total leptonic invariant mass $q^2$ and 
the hadronic invariant mass $M_X^2$.  No experimental cuts are applied.
Our analytic results are obtained via an asymptotic expansion 
around the limit $m_b \simeq m_c$. 
After converting the scheme for the bottom mass to the kinetic scheme
we compare the size of higher QCD corrections to the contributions
from $1/m_b^2$ and $1/m_b^3$ power corrections and to the relative uncertainties.

\bigskip


\thispagestyle{empty}

\newpage


\section{\label{sec::intro}Introduction}

Semileptonic $B$-meson decays mediated by the $b \to c \ell \bar \nu_\ell$ transition 
are sensitive to the absolute value of the Cabibbo-Kobayashi-Maskawa (CKM) matrix element $V_{cb}$. 
In the last years, measurements from BABAR, Belle and LHCb showed a puzzling discrepancy
of about 3 standard deviations between the determinations of $|V_{cb}|$ from 
exclusive and inclusive decays~\cite{HFLAV:2019otj}. 
A simultaneous resolution of the  $|V_{cb}|$ (and $|V_{ub}|$) discrepancy is hardly 
possible in term of new physics~\cite{Crivellin:2014zpa}.
Thus, further scrutiny of theoretical and experimental analyses are needed in order to shed light on the puzzle.

In this paper we focus on higher order QCD corrections to the kinematic moments of
inclusive semileptonic  $B\to X_c \ell \bar \nu_\ell$ decays.
The theory underlying inclusive decays is based on a local operator product expansion,
the Heavy Quark Expansion (HQE)~\cite{Manohar:2000dt,Benson:2003kp,Bauer:2004ve,Gambino:2013rza},
which allows to predict sufficiently inclusive decay observables, as the total semileptonic rate
or moments of kinematic spectra, as an expansion in inverse powers of the bottom quark mass.
In a first approximation, the process can be described as free quark decay.
Bound-state and hadronization effects are incorporated in a set of physical HQE parameters which
appear starting at order $1/m_b^2$.

Inclusive kinematic distributions represent a portal to a precise
determination of the HQE parameters and $|V_{cb}|$.  Lepton energy moments and
moments of hadronic invariant mass have been extensively measured at $B$
factories and their prediction is know up to next-to-next-to leading order
(NNLO) for free
quarks~\cite{Pak:2008cp,Melnikov:2008qs,Biswas:2009rb,Gambino:2011cq}, and
next-to-leading order (NLO) at order
$1/m_b^2$~\cite{Becher:2007tk,Alberti:2012dn,Alberti:2013kxa}.  Moments of the
leptonic invariant mass have also received attention in the recent years due
to their dependence on a smaller set of HQE parameters~\cite{Fael:2018vsp}.
Results for the NLO corrections up to $1/m_b^3$ have been presented
in~\cite{Mannel:2021zzr}.

It is the aim of this paper to compute the next-to-next-to-next-to-leading
order (N$^3$LO) corrections of kinematic moments and assess their relevance
for the global fits of $|V_{cb}|$.  Recently, we presented the N$^3$LO
corrections to the semileptonic width~\cite{Fael:2020tow} and the relation
between the on-shell and kinetic mass of the bottom
quark~\cite{Fael:2020iea,Fael:2020njb}.  In these works we took advantage of
the heavy daughter expansion~\cite{Dowling:2008mc} to determine finite charm
mass effects via an asymptotic expansion in the parameter
$\delta = 1 -m_c/m_b$, where $m_c$ and $m_b$ are the charm and bottom masses,
respectively.  A similar strategy can be applied to compute moments in case no
experimental cuts are applied, i.e.\ moments of kinematic distributions
integrated over the whole phase space.  We present in this work the first four
moments of the charged-lepton energy $E_\ell$, the total leptonic invariant
mass $q^2$ and the hadronic invariant mass $M_X^2$.  We study the behaviour of
the perturbative series in the so-called kinetic scheme, in which the moments
are expressed in terms of the kinetic mass of the bottom quark
mass~\cite{Bigi:1996si,Czarnecki:1997sz,Fael:2020iea,Fael:2020njb}.
Furthermore we estimate the theory uncertainty due to the finite expansion
depth in $\delta$.

We aim at validating the theoretical uncertainty estimates entering the
$|V_{cb}|$ extraction and at identifying the precision level below which
N$^3$LO corrections need to be taken into account.  Usually, kinematic moments
are measured with various kind of lower cuts on $E_\ell$ or $q^2$.  On the one
hand these cuts suppress background from low-energy electrons.  On the other
hand measurements with different cut values provide extra information on the
HQE parameters.  For a prediction of such kind of observables it is necessary to
compute the differential rate to third order.

The paper is organized as follows. In Sec.~\ref{sec::details} we introduce the
notation and present technical details of the calculation of the moments and
also of the total rate presented in Ref.~\cite{Fael:2020tow}.  We discuss in
Sec.~\ref{sec::momOS} the numerical results in the on-shell scheme and discuss
the theoretical uncertainties due to the finite expansion in the parameter
$\delta$.  Numerical results in the kinetic scheme are given in
Sec.~\ref{sec:momKIN}. NLO corrections to the power-suppressed terms of the
$q^2$ moments are considered in Sec.~\ref{sec:AvsB} and in Sec.~\ref{sec:conclusions}
we draw our conclusions. In the Appendix we collect convenient formulae for
one-loop integrals with arbitrary tensor rank and analytic expressions for
the power-suppressed $q^2$ moments including perturbative one-loop corrections.



\section{Details of the calculation}
\label{sec::details} 

\subsection{Moment definitions}
We consider in perturbative QCD the inclusive decay of a bottom quark
\begin{equation}
  b(p) \to X_c (p_x) \ell (p_\ell) \bar \nu_\ell (p_\nu),
\end{equation}
where $X_c$ generically denotes a state containing a charm quark, 
plus additional gluons and/or quarks.
In the rest frame of the bottom quark we have $p = (m_b,\vec 0)$.
Leptons are considered to be massless.
We denote the momentum of the lepton pair by $q= p_\ell + p_\nu$ and 
the total momentum of the hadronic system by $p_x = p-q$.
In the following we study moments of the invariant mass $q^2$, the hadronic invariant mass $M_X^2$
and the charged-lepton energy $E_\ell$.
Moreover, quantities denoted by ``$\hat{\phantom a}$'' refer to dimensionless quantities, 
normalized to the $b$ quark mass, e.g.\ $\hat q^2 = q^2/m_b^2, \hat E_\ell = E_\ell/m_b$.

We compute moments of the differential rate where no restriction is applied on the final state particles.
For their calculation we use the optical theorem in analogy to Ref.~\cite{Fael:2020tow} where the semileptonic width was presented.
As building blocks it is convenient to introduce in the bottom quark rest frame the moments of the leptonic energy $q_0 = p \cdot q/m_b$
 and the leptonic invariant mass $q^2$,
\begin{equation}
  Q_{i,j} = 
  \frac{1}{\Gamma_0}
  \int 
  dE_\ell \, dq_0 \, d q^2 \,
  (q^2)^i (q_0)^j
  \frac{d^3 \Gamma}{dE_\ell \, dq_0 \, d q^2},
  \label{eqn:QijDefinition}
\end{equation}
and moments of the charged-lepton energy $E_\ell = p_\ell \cdot p/m_b$
\begin{equation}
  L_{i} = 
  \frac{1}{\Gamma_0}
  \int 
  dE_\ell \, dq_0 \, d q^2\,
  (E_\ell)^i
  \frac{d^3 \Gamma}{dE_\ell \, dq_0 \, d q^2}
  \label{eqn:LiDefinition}
\end{equation}
with the normalization factor
\begin{equation}
  \Gamma_0 = \frac{m_b^5 G_F^2 |V_{cb}|^2}{192 \pi^3}.
\end{equation}
Note that $Q_{0,0} = L_0$ corresponds to the total semileptonic rate computed in~\cite{Fael:2020tow} (divided by $\Gamma_0$).
Moments are written as a series expansion in the strong coupling constant $\alpha_s(\mu_s)$,
\begin{align}
  Q_{i,j}&= \sum_{n\ge 0} Q_{i,j}^{(n)} \left( \frac{\alpha_s(\mu_s)}{\pi} \right)^n, &
  L_{i} &= \sum_{n\ge 0} L_i^{(n)} \left( \frac{\alpha_s(\mu_s)}{\pi} \right)^n.
  \label{eqn:PerturbativeExpansionMoments}
\end{align}
Normalized moments are defined by
\begin{align}
  \langle (q^2)^n \rangle &\equiv
  \frac{Q_{n,0}}{Q_{0,0}}, &
  \langle E_\ell^n \rangle &\equiv
  \frac{L_{n}}{L_{0}}\,,
  \label{eqn:NormalizedMomentsDefinition}
\end{align}
with $n\ge1$ and centralized moments are given by
\begin{align}
  q_1 &\equiv \langle q^2 \rangle, &
  q_n &\equiv \left\langle \left( q^2 - \langle q^2 \rangle\right)^n \right\rangle , \notag \\
  \ell_1 &\equiv \langle E_\ell \rangle, &
  \ell_n &\equiv \left\langle \left( E_\ell -  \langle E_\ell \rangle \right)^n \right\rangle\,,
  \label{eqn:CentralizedMomentsDefinition}
\end{align}
where $n\ge2$. Predictions for normalized and centralized moments can be
obtained by inserting the perturbative expansions
\eqref{eqn:PerturbativeExpansionMoments}
into~\eqref{eqn:NormalizedMomentsDefinition} or
\eqref{eqn:CentralizedMomentsDefinition} and re-expanding in $\alpha_s$.

The hadronic invariant mass is related to parton level quantities by
\begin{equation}
  M_X^2 \equiv (p_B-q)^2 = M_B^2-2 M_B q_0+q^2, 
\end{equation}
where $p_B$ and $M_B$ are the momentum and the mass of the $B$ meson, respectively.
We assume that the bottom quark and the $B$ meson have the same velocity, i.e.\
$p_B^\mu = M_B v^\mu$ and $p = m_b v^\mu$.
The moments of $M_X$ are given by linear combinations of the $Q_{i,j}$ moments:
\begin{align}
  M_n &=  
  \frac{1}{\Gamma_0}
  \int 
  dE_\ell \, dq_0 \, d q^2 \,
  (M_B^2-2 M_B q_0+q^2)^n
  \frac{d^3 \Gamma}{dE_\ell \, dq_0 \, d q^2}  \notag \\
  &=
  \sum_{i=0}^n \sum_{j=0}^i
  \binom{n}{i}
  \binom{i}{j}
  (M_B^2)^{n-i}(-2M_B)^{i-j}
  Q_{j,i-j} \, .
  \label{eqn:MnDefinition}
\end{align}
Normalized and centralized moments are defined as
\begin{align}
  \langle (M_X^2)^n \rangle &\equiv
  \frac{M_n}{M_0}, &
  h_1 &\equiv \langle M_X^2 \rangle, &
  h_n &\equiv \left\langle 
  \left( M_X^2 - \langle M_X^2 \rangle\right)^n \right\rangle. 
        \label{eq::MnDefinition_norm}
\end{align}

\subsection{\boldmath Asymptotic expansion}

\begin{figure}[t]
  \centering
  \begin{tabular}{ccc}
    \raisebox{0em}{\includegraphics[width=0.25\textwidth]{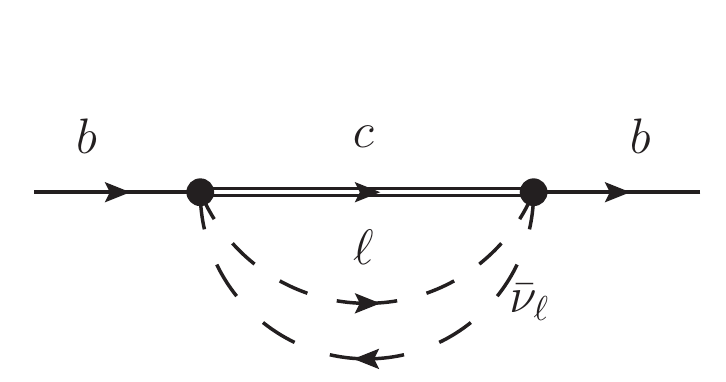}} &
    \raisebox{0em}{ \includegraphics[width=0.25\textwidth]{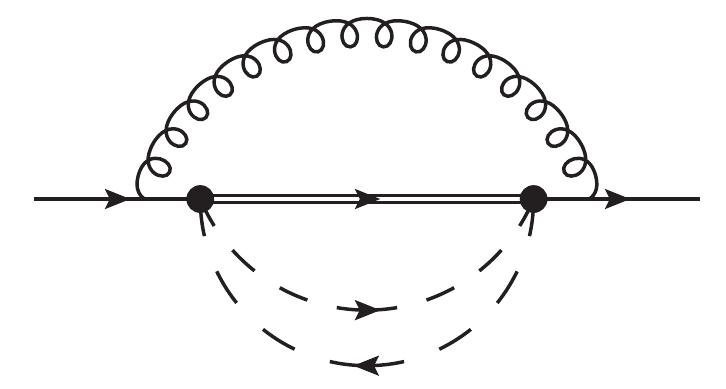}} &
    \includegraphics[width=0.25\textwidth]{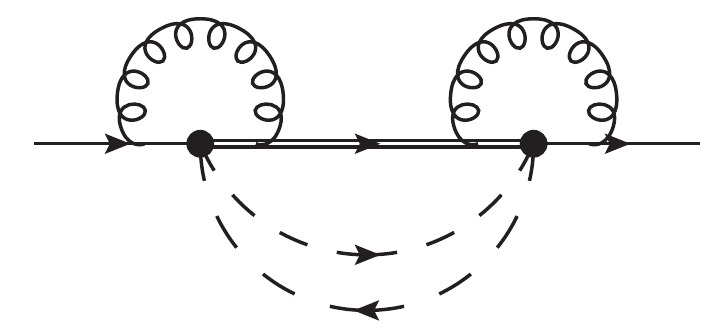} \\
    (a) & (b) & (c) \\[10pt]
    \raisebox{0em}{\includegraphics[width=0.25\textwidth]{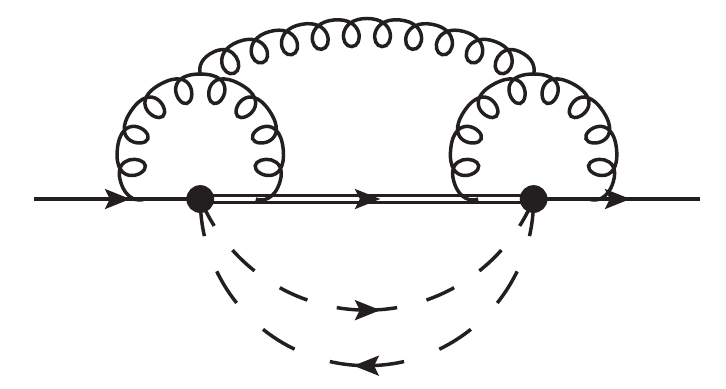}} &
    \raisebox{0em}{\includegraphics[width=0.25\textwidth]{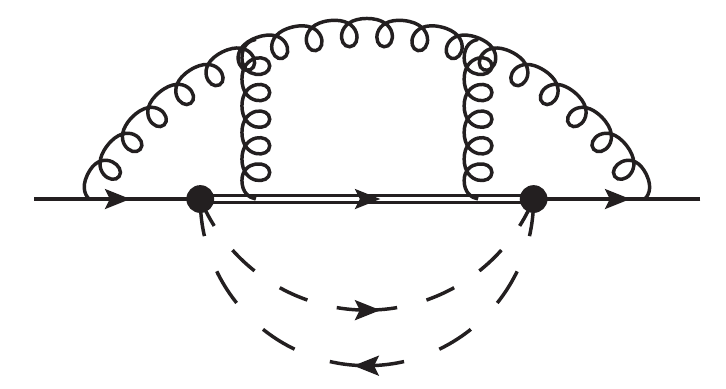}} &
    \includegraphics[width=0.25\textwidth]{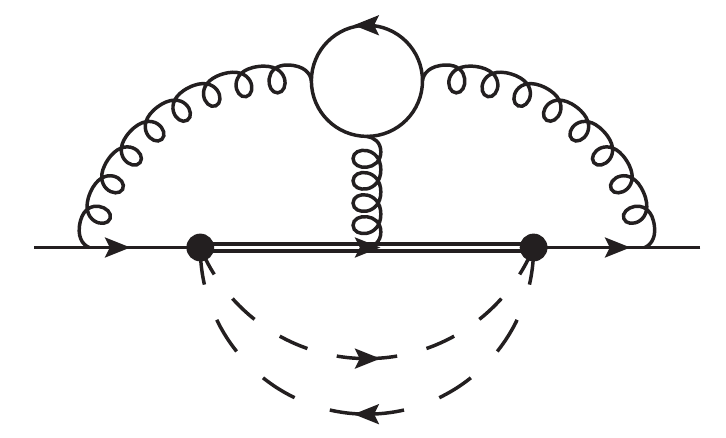} \\
    (d) & (e) & (f) \\
  \end{tabular}
  \caption{\label{fig:diag}
    Sample Feynman diagrams which contribute to the
    forward scattering amplitude of a bottom quark at LO (a), NLO (b), NNLO
    (c) and N$^3$LO (d-f). Straight, curly and dashed lines represent quarks,
    gluons and leptons, respectively. The weak interaction mediated by the $W$
    boson is shown as a black dot.}
\end{figure}

Let us now describe the calculation of $Q_{i,j}$ and $L_i$.  With the help of
the optical theorem we can express the $b \to X_c \ell \bar \nu_\ell$ matrix
element integrated over the whole phase space
in~Eqs.~\eqref{eqn:QijDefinition} and \eqref{eqn:LiDefinition} in terms of the
discontinuity of the $b \to b$ forward scattering amplitude (for sample
Feynman diagrams see Fig.~\ref{fig:diag}).  Moments without cuts are simply
obtained by multiplying the forward scattering amplitude by the weight
function $(q^2)^i (q\cdot v)^j$ or $(p_\ell \cdot v)^i$ for the $Q_{i,j}$ and
$L_i$, respectively.  The leading order prediction is obtained from the
two-loop diagram in Fig.~\ref{fig:diag}(a) where the internal lines correspond
to the neutrino, the charged lepton and the charm quark.  The weak interaction
is shown as an effective vertex.  To compute QCD corrections up to
$O(\alpha_s^3)$ we have to add up to three more loops (see
Fig.~\ref{fig:diag}(b) to (f)).

An exact computation of five-loop diagrams with two mass scales ($m_b$ and $m_c$) is out 
of range using current methods.
We obtain finite charm mass effects by performing an asymptotic expansion 
in the parameter $\delta = 1-m_c/m_b \ll 1$, 
i.e.\ we expand the Feynman diagrams around the equal mass limit $m_c \simeq m_b$, 
which we realize with the method of regions~\cite{Beneke:1997zp,Smirnov:2012gma}. 
We call this approach the \textit{$\delta$-expansion}. 
The opposite limit $\rho = m_c/m_b \ll 1$ (the \textit{$\rho$-expansion}) was adopted 
in~\cite{Pak:2008cp} for the evaluation of the width to $O(\alpha_s^2)$.

It has been shown that the $\delta$-expansion converges quite fast for the physical values of
quark masses $ \delta \simeq 0.7$~\cite{Dowling:2008mc,Fael:2020tow,Czakon:2021ybq}. 
Moreover compared to an expansion around the opposite limit ($\rho \simeq 0.3$), 
the $\delta$-expansion offers two crucial advantages:
\begin{enumerate}
  \item The number of regions to be calculated is considerably smaller.
  \item The $\delta$-expansion yields a factorization of the multi-loop 
    integrals which allows us to integrate at least two loop momenta without applying
    integration-by-part (IBP) relations. 
    A computation up to $O(\alpha_s^n)$ becomes a $n$-loop problem, 
    even if we start with $(n+2)$-loop Feynman diagrams.  
\end{enumerate}

\begin{figure}[h]
  \centering
  \includegraphics[width=0.5\textwidth]{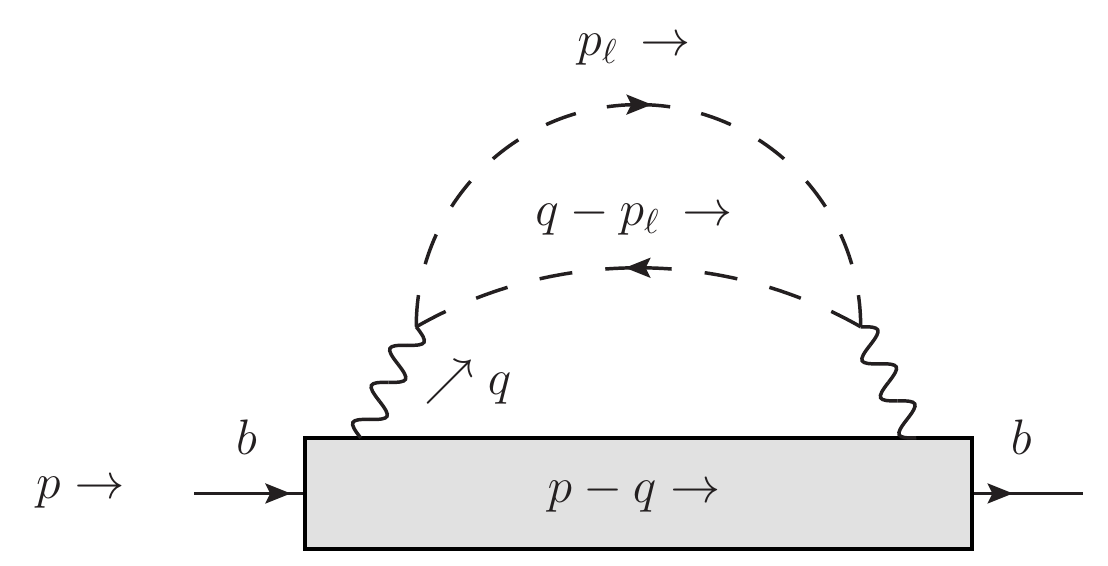}
  \caption{Our convention for the loop momentum routing.  Charged lepton and
    neutrino momenta are $p_\ell$ and $q-p_\ell$, respectively.  The external
    bottom quark momentum is $p$.  Additional loops of gluons and quarks are
    denoted generically with the gray blob.  The arrows on the fermion lines
    indicate the fermion direction whereas the arrows next to the lines denote
    the momentum flow.}
  \label{fig:MomentumRouting}
\end{figure}

In the following we elaborate on these two points.
It is convenient to route the bottom quark momentum $p$ along the external fermion line
and we chose the momentum routing in the lepton-neutrino loop as shown in Fig.~~\ref{fig:MomentumRouting}.
Then the loop integrals w.r.t.\ $p_\ell$ take the form
\begin{equation}
  I_1^{\mu_1 \dots \mu_N}(d,q^2;n_1,n_2) =
  \int \frac{d^d p_\ell}{(2\pi)^d}
  \frac{p_\ell^{\mu_1} \dots p_\ell^{\mu_N}}
  {(-p_\ell^2)^{n_1} (-(p_\ell-q)^2)^{n_2}}, 
\end{equation}
where $n_1$ and $n_2$ are integers and $d= 4-2\epsilon$ is the space-time dimension.
For such integrals one can derive a closed formula for arbitrary tensor rank $N$ 
(see e.g.~\cite{Smirnov:2012gma} and Eq.~\eqref{eqn:tdecmassless} in Appendix~\ref{app::tensor}). 
After performing the $p_\ell$ integration, we obtain integrals with an effective propagator $1/q^2$
raised to an $\epsilon$-dependent power. 

Next we apply the  method  of regions to construct the $\delta$-expansion.
There are only two possible scalings for each loop momentum $k$~\cite{Smirnov:2012gma}:
\begin{itemize}
  \item \emph{hard} (h): $|k^\mu| \sim m_b$,
  \item \emph{ultra-soft} (u): $|k^\mu| \sim \delta \cdot m_b = m_b-m_c$.
\end{itemize}
We choose the notion ``ultra-soft'' for the second scaling in analogy to the
calculation of the relation between the pole and the kinetic mass of a heavy
quark, see~\cite{Fael:2020iea,Fael:2020njb}.  For all diagrams, we checked
with the program \texttt{asy.m}~\cite{Pak:2010pt} that a naive scaling
assignment to the individual loop momenta correctly identifies all relevant
regions.

Since there is only one scale in the problem, the parameter $\delta$ (we set
$m_b=1$), an imaginary part arises only through the appearance of
$\log(-\delta)$, i.e.\ only if $\delta$ appears in the denominator of one of
the charm propagators.  This implies that the combination $k-q$, where $k$ is
a loop momentum running through a charm quark line and $q=p_\ell+p_\nu$, must
be ultra-soft for at least one of the charm propagators. Otherwise no imaginary
part arises.  Furthermore, the momentum $q$ of the lepton pair always has to
scale ultra-soft which means that all regions where $q$ scales hard are
discarded.  To clarify this point, let us consider for instance the following
propagator:
\begin{equation}
  \frac{1}{(p-q+k)^2 - m_c^2} =
  \frac{1}{ 2p\cdot (k-q) + (k-q)^2 + 2 \delta - \delta^2  },
\end{equation}
where $k$ denotes some generic linear combination of loop momenta other than $q$.
If $k-q$ scales hard ($p$ is considered always hard), we expand the charm propagators as follows
\begin{equation}
  \frac{1}{(p-q+k)^2 - m_c^2} \overset{\footnotesize \text{h}}{=}
  \frac{1}{ 2p\cdot (k-q) + (k-q)^2} + O(\delta) \,.
\end{equation}
Thus, no $\delta$ is left in the denominator and no imaginary part appears. 
If $k-q$ is ultra-soft we have
\begin{equation}
  \frac{1}{(p-q+k)^2 - m_c^2} \overset{\footnotesize \text{u}}{=}
  \frac{1}{ 2p\cdot (k-q) + 2 \delta } + O(\delta^0) .
  \label{eq::charm_prop}
\end{equation}
After integration the $\delta$ in  the denominator yields a $\log(-\delta)$ term and thus an imaginary part.

At this point we exploit the fact that $q$ is always ultra-soft which allows
us to perform a further integration.  Integrals where the loop momenta are
hard factorize from the integration w.r.t.\ $q$.  The crucial observation is
that also in case $q$ and $k$ are both ultra-soft the integrations factorize.
In fact, having chosen the momentum routing as in
Fig.~\ref{fig:MomentumRouting}, the dependence in the charm propagators on $q$
and $\delta$ is always of the form $(-2p\cdot q + 2 \delta)$ as can be seen
from Eq.~(\ref{eq::charm_prop}).  Taking
advantage of the linearity of the charm propagators in the ultra-soft region,
we can pull out the global factor $(-2p\cdot q + 2 \delta)$ from each propagator
by rescaling the loop momenta.  For instance, for the following two-loop
integral we have\footnote{Note that we set $m_b=1$.}
\begin{multline}
  \int  
\frac{d^d q \, d^d k}
{(q^2)^{n_1} (k^2)^{n_2} (2p\cdot k -2 p \cdot q+2 \delta)^{n_3}  } 
\quad \overset{k \to k (-2p\cdot q +2 \delta)}{=}\\
\int \frac{d^d q}{(q^2)^{n_1} (-2 p\cdot q+2\delta)^{-d+2n_2+n_3} }
\times \int \frac{d^dk }{ (k^2)^{n_2} (2p\cdot k +1 )^{n_3}}.
\end{multline}
Thus the $q$ integration also factorizes for ultra-soft loop momentum $k$ and 
therefore we can always evaluate the $q$-integration independently on the other loop momenta. 
The tensor integrals 
\begin{equation}
  I_2^{\mu_1 \dots \mu_N}(d,\delta,p^2; n_1,n_2) =
  \int \frac{ d^d q}{(2\pi)^d}
  \frac{q^{\mu_1} \dots q^{\mu_N}}
  {(-q^2)^{n_1} (-2 p \cdot q + 2 \delta)^{n_2}},
\end{equation}
can be directly evaluated using Eq.~\eqref{eqn:tdecultrasoft} in Appendix~\ref{app::tensor}.
In conclusion, we are able to analytically carry out the integration w.r.t.\ $p_e$ and $q$
without the need of an IBP reduction and we remain with $n$ momentum integrations at order
$\alpha_s^n$. Each of these momenta can either be hard or ultra-soft.

With the same approach, it is possible to integrate all one-loop hard or ultra-soft contributions
which leaves purely hard or ultra-soft integrals at two and three loops.
We reduce them to master integrals via standard IBP reduction.
We summarize all regions at order $\alpha_s,\alpha_s^2$ and $\alpha_s^3$ in Tab.~\ref{tab:regions}.
Those labeled in red required an IBP reduction, while the other regions factorize 
and are computed with the help of Eqs.~(\ref{eqn:tdecmassless}) to (\ref{eqn:tdecultrasoft}).

After asymptotic expansion of the Feynman integrals one gets linearly
dependent propagators. It is thus necessary to perform a partial fraction
decomposition in order to arrive at proper input expressions for the IBP reduction.
The methods employed for the partial fraction decomposition and the mappings among different integral
families closely follow those described in Ref.~\cite{Fael:2020njb},
in particular we used the program \texttt{LIMIT}~\cite{Herren:2020ccq} to automate
the partial fraction decomposition in case of linearly dependent denominators.
For all cases where at least one of the regions is ultra-soft we can take
over the master integrals from~\cite{Fael:2020iea,Fael:2020njb}.  
For some of the (complicated) three-loop triple-ultra-soft master integrals, higher order $\epsilon$ terms are needed. 
The method used for their calculation and the results are given Ref.~\cite{Fael:2020njb}. 
All triple-hard master integrals can be found in Ref.~\cite{Lee:2010ik}.
\begin{table}
  \centering
  \begin{tabular}{cc}
    order & regions \\
    \hline
    $\alpha_s$ &  u, h\\
    $\alpha_s^2$ & \textbf{\color{red} uu}, \textbf{ \color{red}hh}, hu \\
    $\alpha_s^3$ & \textbf{\color{red}uuu}, \textbf{\color{red}hhh}, h\textbf{\color{red}uu}, u\textbf{\color{red}hh}
  \end{tabular}
  \caption{Relevant regions for the loop momenta $k_1,k_2,k_3$ up to $O(\alpha_s^3)$: hard (h) and ultra-soft (u). 
    Regions written in black factorize, leaving at most two- or three-loop integrals (in red) to be reduce by means of IBP relations.}
    \label{tab:regions}
\end{table}

For all moments we have computed the first 16, 11 and 8 terms in the $\delta$-expansion
at order $\alpha_s, \alpha_s^2$ and $\alpha_s^3$, respectively.
Note that the leading power of $\delta$ is different for each moment:
\begin{align}
  \text{leading power of $\delta$ for } Q_{i,j}: &\quad \delta^{5+2i+j}, \notag \\
  \text{leading power of $\delta$ for } L_{i}: &\quad \delta^{5+i}.
\end{align}
This means for example that the $\alpha_s^3$ correction to the width is computed up to order $\delta^{12}$, 
while for the third lepton energy moment $L_3$ the expansion extends to $\delta^{15}$.
Note that the leading term for the latter is $\delta^8$. 

The chosen expansion depths are a compromise between precision of our prediction and computational resources.
To achieve sufficient precision, especially for the centralized moments (see next section),
we had to perform a deep expansion in $\delta$ of the Feynman propagators, 
up to 8th or 10th order which has led to intermediate expressions of the order of 100~GB for each diagram.
They must  be handled carefully by \texttt{FORM}~\cite{Ruijl:2017dtg} in order to avoid 
an explosion of the number of terms.  

Furthermore for some of the integral families, individual propagators are raised
to positive and negative powers up to 12, 
which  constitute a non-trivial task for the IBP reduction programs. 
The latter could be handled thanks to a private version of {\tt FIRE}~\cite{Smirnov:2019qkx} 
combined with {\tt  LiteRed}~\cite{Lee:2012cn}.
For the subset of integrals which are needed for the expansion up to $\delta^{10}$ we also 
use the stand-alone version of {\tt   LiteRed} as a cross-check.

There is an additional complication in the computation of the charged-lepton
energy moments.  They are computed by introducing the factor
$(p_\ell \cdot v)^i$ in the integrand of the electron-neutrino loop, which
make them dependent on the parity-odd part of the amplitude.  As a consequence
the traces which contain an odd number of $\gamma_5$ matrices does not cancel
anymore and we have to deal with traces involving $\gamma_5$ in $d$ dimension.
We adopt the so-called Larin prescription~\cite{Larin:1993tq} and substitute
\begin{eqnarray}
  \gamma^\mu \gamma^5 &\to& 
  \frac{i}{3!} \, \varepsilon^{\mu\nu\rho\sigma} 
  \frac{\left( 
    \gamma_{\nu}\gamma_\rho\gamma_{\sigma} - 
    \gamma_{\sigma}\gamma_\rho\gamma_{\nu}
  \right)}{2}
  \, ,
  \label{eq::ax-ve}
\end{eqnarray}
in those cases where one instance of axial-vector current is present in a 
leptonic trace and one in the bottom-charm fermion line.
After evaluating the traces of $\gamma$ matrices,
we contract the two Levi-Civita tensors 
and interpret the result in $d$ dimensions. 
In case two $\gamma_5$ matrices are present in a trace, we simply anti-commute $\gamma^5$.

For the contributions where the Larin prescription have been used, an
additional $\overline{\mathrm{MS}}$ renormalization constant has to be taken
into account.  An axial-vector current treated with the Larin prescription
must be renormalized with the factor~\cite{Larin:1991tj,Larin:1993tq}
\begin{align}
    Z_A &= 1
    + \left(\frac{{\alpha_s}}{\pi}\right)^2
    \frac{1}{\epsilon} 
    \biggl(
        \frac{11}{24} C_A C_F
        -\frac{1}{6} C_F T_F n_f
    \biggr)
    + \left(\frac{{\alpha_s}}{\pi}\right)^3 
    \biggl[
        \frac{1}{\epsilon^2} 
        \biggl(
            -\frac{121}{432} C_A^2 C_F
            +\frac{11}{54} C_A C_F T_F n_f
            \nonumber \\ &
            -\frac{1}{27} C_F T_F^2 n_f^2
        \biggr)
        +\frac{1}{\epsilon} 
        \biggl(
            \frac{1789}{2592} C_A^2 C_F
            -\frac{77}{144} C_A C_F^2
            -\frac{26}{81} C_A C_F n_f T_F
            +\frac{1}{9} C_F^2 T_F n_f 
            \nonumber \\ &
            +\frac{1}{162} C_F T_F^2 n_f^2
        \biggr)
    \biggr]
    ,
\end{align}
where $T_F=1/2$ is the trace normalization and
$C_F=4/3$ and $C_A=3$ are the Casimir operators of the fundamental and
the adjoint representation of $SU(3)$, respectively,
$\alpha_s \equiv \alpha_s^{(n_f)}(\mu_s)$, $n_f$ is the number of active flavours and $\mu_s$ 
is the renormalization scale of the coupling constant.
Furthermore one has to introduce a finite renormalization constant 
in order to restore the correct Ward identity:
\begin{align}
    Z_5 &=
    1
    - \frac{{\alpha_s}}{\pi} C_F
    +\left(\frac{{\alpha_s}}{\pi}\right)^2
    \biggl(
            -\frac{107}{144} C_A C_F
            +\frac{11}{8} C_F^2
            +\frac{1}{36} C_F T_F n_f
    \biggr)
    \nonumber \\ &
    +\left(\frac{{\alpha_s}}{\pi}\right)^3 
    \biggl[
            C_A^2 C_F
            \biggl(
                    -\frac{2147}{1728}
                    +\frac{7 \zeta_3}{8}
            \biggr) 
            + C_A C_F^2
            \biggl(
                    \frac{2917}{864}
                    -\frac{5 \zeta_3}{2}
            \biggr) 
            +C_F^3 
            \biggl(
                    -\frac{185}{96}
                    +\frac{3 \zeta_3}{2}
            \biggr) 
            \nonumber \\ &
            +C_A C_F T_F n_f 
            \biggl(
                    \frac{89}{648}
                    +\frac{\zeta_3}{3}
            \biggr) 
            +C_F^2 T_F n_f 
            \biggl(
                    -\frac{31}{432}
                    -\frac{\zeta_3}{3}
            \biggr) 
            +\frac{13}{324} C_F T_F^2 n_f^2 
    \biggr]
    .
\end{align}

Finally, it is interesting to note that the natural expansion parameter
arising from the Feynman diagrams is actually $\delta' = 1-m_c^2/m_b^2$ as odd
powers of $m_c$ do not appear in the differential rate because of the $V$-$A$
weak interaction~\cite{Berman1962SomeCO,Roos:1971mj}.  Odd powers of $m_c$ can
appear in the lepton energy moments at intermediate steps when using the Larin
scheme.  In particular, they are present in the higher $\epsilon$ terms
of the lower-order corrections. In this case we rewrite
$m_c^{2n+1} = m_c (1-\delta')^n m_b^{2n}$ and treat $m_c$ as additional
parameter.  However, after renormalization, we verify that all odd powers of
$m_c$ vanish.

The use of $\delta'$ further reduces the size of intermediate expressions.
Only at the very end, after renormalization, we re-express our results in term of 
$\delta=1-m_c/m_b=1-\sqrt{1-\delta^'}$ since the series in $\delta$ converges faster.
This fact can be understood by comparing, for instance, the behaviour of the 
tree level decay rate which is proportional to the function
\begin{equation}
  f(\rho) = 1 - 8 \rho^2 + 8 \rho^6 - \rho^8 - 12 \rho^4 \log(\rho^2),
  \label{eq::fLO}
\end{equation}
with $\rho=m_c/m_b$. If we substitute $\rho=1-\delta$, at higher orders
in $\delta$ the series is governed by the expansion of $\rho^4 \log(\rho^2)$
which is given by
\begin{align}
  \rho^4 \log(\rho^2) &= 
  - 2 (1-\delta)^4 \log(1-\delta) =
  - 2 (1-\delta)^4 \sum_{m= 1}^{\infty} \frac{\delta^m}{m} \notag \\
  &=
  -2 \delta +7 \delta^2 -\frac{26}{3} \delta^3 +\frac{25}{6} \delta^4
  -\sum_{n = 5}^{\infty}
  \frac{48}{n (n-1) (n-2) (n-3) (n-4) } \delta^n.
\end{align}
Instead, if we substitute $\rho^2=1-\delta'$ we obtain
\begin{align}
  \rho^4 \log(\rho^2) &=
  (1-\delta')^2 \log(1-\delta') = 
  -(1-\delta')^2 \sum_{m=1}^{\infty} \frac{ (\delta^')^m}{m} \notag \\
  &= - \delta'+\frac{3}{2} (\delta^')^2 -
  \sum_{n=3}^{\infty}
  \frac{2}{n(n-1)(n-2)}  (\delta^')^n
    \,.
\end{align}
If we adopt $\delta$ as expansion parameter, the coefficients in the series are suppressed by $1/n^5$ for large $n$,
while for $\delta'$ the coefficients are suppressed only by $1/n^3$. 
This fact suggest to use $\delta$ as expansion parameter also in the prediction at higher orders in $\alpha_s$.



\section{\boldmath Results in the on-shell scheme}
\label{sec::momOS} 

Our main results are analytic expressions for the moments $Q_{i,j}$ and $L_i$,
with $i+j \le 4$, which can be downloaded from~\cite{progdata}.  In this
section we first assess the uncertainty of the central moments related to the
$\delta$-expansion.  In the next section we convert our results to the kinetic
scheme and compare the size of the $O(\alpha_s^3)$ terms to experimental
results and to the size of higher power corrections.

Let us fix for the numerical evaluation $m_b^{\mathrm{OS}} = 4.6 $ GeV and
$m_c^{\mathrm{OS}} = 1.15$ GeV which leads to $\delta = 0.75$.  We use
$M_B = 5.279 $ GeV for the $M_X^2$ moments and set the renormalization scale
$\mu_s = m_b^{\mathrm{OS}} $.  The $\delta$-expansion provides precise
approximations for $Q_{i,j}$ and $L_i$.  To give an idea of the convergence,
we show the size of the different terms in the series at order $\alpha_s^3$
for three selected moments:
\begin{align}
  \hat Q_{0,0}^{(3)} &=
  -44.9615(1\dflag{5} - 0.527532 \dflag{6} + 4.38372 \dflag{7} - 2.54593 \dflag{8} + 0.102771 \dflag{9} 
  \notag \\
&\quad \quad + 0.0168158 \dflag{10} + 0.00263043 \dflag{11} + 0.00216016 \dflag{12}),
    \notag \\
  \hat Q_{4,0}^{(3)} &=
    -0.703488(1\dflag{13} - 0.527532 \dflag{14} + 2.79417 \dflag{15} - 1.488 \dflag{16} - 0.077824 \dflag{17} 
      \notag \\
   & \quad \quad - 0.0329351 \dflag{18} - 0.0139737 \dflag{19} - 0.0058596 \dflag{20}), \notag \\
\hat L_1^{(3)} &=
      -16.8605(1\dflag{6} - 0.992521 \dflag{7} + 5.56695 \dflag{8} - 4.14032 \dflag{9} + 0.754176 \dflag{10} 
      \notag \\
    &\quad \quad - 0.0251885 \dflag{11} - 0.0103673 \dflag{12} - 0.00171797 \dflag{13}),
\end{align}
where the subscripts are introduced to flag the different terms in the $\delta$-expansion.
The first equation corresponds to the expansion of the rate in~\cite{Fael:2020tow}.
We observe that at $O(\alpha_s^3)$ the  precision reached with eight terms is well below the relative $1\%$ level.

However, the accuracy on the centralized moments reduces.  To compute
centralized moments, we insert the analytic results of $L_i^{(n)}$ or
$Q_{i,j}^{(n)}$ in Eqs.~\eqref{eqn:CentralizedMomentsDefinition}
and~\eqref{eq::MnDefinition_norm} and re-expand in $\alpha_s$ to third order.  The
re-expansion in $\alpha_s$ of numerator and denominator is subject to strong
cancellations.  We do not re-expand in $\delta$.  The correction to
centralized moments at order $\alpha_s^n$ involves non-trivial combinations of
the moments $L_i^{(m)}$ or $Q_{i,j}^{(m)}$, where $m$ ranges from $0$ to $n$.
A simple re-expansion in $\delta$, let us say up to the eighth term at order
$\alpha_s^3$, spoils the delicate cancellations happening among different
moments with $m < n$, which are actually computed to higher precision in
$\delta$.  Therefore we suggest not to re-expand in $\delta$ quantities
derived from $L_i^{(m)}$ or $Q_{i,j}^{(m)}$ since they represent the best
possible approximation.

We estimate the final accuracy in the following way.  We consider the moments
with the highest computed term in $\delta$ and insert numerical values for the
masses.  Then we re-evaluate each moment removing the last term in the
$\delta$-expansion at each order in $\alpha_s$.  The difference between these
two numerical predictions is quoted as uncertainty.

For the centralized $q^2$ moments normalized to $m_b$ and expressed in the
on-shell scheme we obtain
\begin{align} 
 \hat q_1 &= &0.218482&\Big[1+0.127423\frac{\alpha_s}{\pi} &&+0.4369(30)\left( \frac{\alpha_s}{\pi} \right)^2 &&-5.34(30)\left( \frac{\alpha_s}{\pi} \right)^3 \Big], \notag \\ 
 \hat q_2 &= &0.0203994&\Big[1+0.138093\frac{\alpha_s}{\pi} &&+0.91584(89)\left( \frac{\alpha_s}{\pi} \right)^2 &&+3.52(33)\left( \frac{\alpha_s}{\pi} \right)^3 \Big], \notag \\ 
 \hat q_3 &= &0.00110423&\Big[1-0.226532\frac{\alpha_s}{\pi} &&+1.137(14)\left( \frac{\alpha_s}{\pi} \right)^2 &&+53.37(59)\left( \frac{\alpha_s}{\pi} \right)^3 \Big], \notag \\ 
 \hat q_4 &= &0.000889517&\Big[1+0.167677\frac{\alpha_s}{\pi} &&+1.5921(11)\left( \frac{\alpha_s}{\pi} \right)^2 &&+15.24(35)\left( \frac{\alpha_s}{\pi} \right)^3 \Big]. 
\end{align}

For the $E_\ell$ moments we find
\begin{align} 
 \hat \ell_1 &= &0.307202&\Big[1-0.0169117\frac{\alpha_s}{\pi} &&-0.6637(30)\left( \frac{\alpha_s}{\pi} \right)^2 &&-15.01(15)\left( \frac{\alpha_s}{\pi} \right)^3 \Big], \notag \\ 
 \hat \ell_2 &= &0.00862693&\Big[1-0.164901\frac{\alpha_s}{\pi} &&-2.0568(59)\left( \frac{\alpha_s}{\pi} \right)^2 &&-35.4(2.9)\left( \frac{\alpha_s}{\pi} \right)^3 \Big], \notag \\ 
 \hat \ell_3 &= &-0.00041875&\Big[1-0.00580025\frac{\alpha_s}{\pi} &&-1.4848(68)\left( \frac{\alpha_s}{\pi} \right)^2 &&-25 (17)\left( \frac{\alpha_s}{\pi} \right)^3 \Big], \notag \\ 
 \hat \ell_4 &= &0.000189369&\Big[1-0.245899\frac{\alpha_s}{\pi} &&-3.534(28)\left( \frac{\alpha_s}{\pi} \right)^2  &&-76 (481)\left( \frac{\alpha_s}{\pi} \right)^3 \Big]. 
\end{align}

For $M_X^2$ moments it is more convenient to normalize the results w.r.t.\ the first
order in $\alpha_s$ since the partonic $X_c$ invariant mass differs from
$m_c$ only starting at $O(\alpha_s)$ due to real radiation. Our results read
\begin{align} 
 \hat h_1 &=& 0.0993848&\Big[2.10166&&+1\frac{\alpha_s}{\pi} &&+14.567(25)\left( \frac{\alpha_s}{\pi} \right)^2 &&+249.0(2.4)\left( \frac{\alpha_s}{\pi} \right)^3 \Big], \notag \\ 
 \hat h_2 &=& 0.0150817&\Big[0.029471&&+1\frac{\alpha_s}{\pi} &&+11.098(59)\left( \frac{\alpha_s}{\pi} \right)^2 &&+152 (40)\left( \frac{\alpha_s}{\pi} \right)^3 \Big], \notag \\ 
 \hat h_3 &=& 0.00342142&\Big[-0.00103783&&+1\frac{\alpha_s}{\pi} &&+9.27(21)\left( \frac{\alpha_s}{\pi} \right)^2 &&+201 (24)\left( \frac{\alpha_s}{\pi} \right)^3 \Big], \notag \\ 
 \hat h_4 &=& 0.001168&\Big[0.000361694&&+1\frac{\alpha_s}{\pi} &&+9.1(1.4)\left( \frac{\alpha_s}{\pi} \right)^2 &&+0 (19) \times 10^3 \left( \frac{\alpha_s}{\pi} \right)^3 \Big]. 
\end{align}

We notice that the centralized  $q_i$ moments are well approximated by the $\delta$-expansion. 
The uncertainties of the $\alpha_s^3$ coefficients are at most of about 10\%.
For the first three $E_\ell$ and $M_X$ centralized moments, we find that our approximation 
is able to determine the size of the third order correction.
However for the moments $\hat \ell_4$ and $\hat h_4$ we observe that our expansion depth
is not deep enough and the large uncertainty is a consequence of severe numerical cancellations.

We noticed also that an uncertainty estimate based on standard error propagation in
general overestimates the uncertainty.  If we assigned to each moment
$L_i^{(m)}$ and $Q_{i,j}^{(m)}$ an error equal to the last known term in
$\delta$ and then combine the uncertainties in an uncorrelated way, we
would find for $\hat q_i$ and $\hat \ell_i$ uncertainties much larger than
those quoted above.  For hadronic moments we would observe errors of similar size.
This fact is likely connected to stronger correlations among the different
expansion terms in $\delta$ for the $q^2$ and $E_\ell$ moments.

We compared our results at ${\cal O}(\alpha_s^2)$ with the values for the $M_X$ and $E_\ell$ moments of
Refs.~\cite{Biswas:2009rb,Gambino:2011cq} and find good agreement.



\section{Transition to the kinetic scheme}
\label{sec:momKIN}

In this section we discuss the impact of higher order QCD corrections once a
short-distance mass scheme is adopted.  Moreover we will compare them to the
power corrections at order $1/m_b^2$ and $1/m_b^3$ to understand the importance
of the $\alpha_s^3$ corrections in the fits for $|V_{cb}|$.

In this work we concentrate on the so called \textit{kinetic scheme} employed
in the fits of Refs.~\cite{Gambino:2013rza,Alberti:2014yda,Bordone:2021oof}.
In this scheme we adopt the kinetic
mass~\cite{Bigi:1996si,Czarnecki:1997sz,Fael:2020iea,Fael:2020njb} for the
bottom quark using
\begin{equation}
m_b^\kin(\mu)= m_b^\OS-
  [\overline \Lambda(\mu)]_\pert
  -\frac{[\mu_\pi^2(\mu)]_\pert}{2m_b^\kin(\mu)} -
  O\left( \frac{1}{(m_b^\kin)^2} \right),
\end{equation}
while the charm quark mass is converted to the  $\overline{\mathrm{MS}}$ scheme.
At the same time, in the kinetic scheme one redefines the heavy-quark-expansion 
parameters $\mu_\pi^2$ and $\rho_D^3$ in the following way:
\begin{align}
  \mu_\pi^2(0) &= \mu_\pi^2(\mu)-[\mu_\pi^2(\mu)]_\pert, &
  \rho_D^3(0) &= \rho_D^3(\mu)-[\rho_D^3(\mu)]_\pert,
  \label{eqn:HQEParametersRedefinition}
\end{align}
where the analytic expressions for $[\overline \Lambda(\mu)]_\pert, \,
[\mu_\pi^2(\mu)]_\pert$ and $[\rho_D^3(\mu)]_\pert$ can be found in the Appendix of Ref.~\cite{Fael:2020njb}.
The Wilsonian cutoff $\mu$ plays the role of scale separation between the short- and long-distance regimes.
We adopt the standard HQE parameter definitions employed in Refs.~\cite{Benson:2003kp,Alberti:2014yda,Bordone:2021oof}:
\begin{align}
  \mu_\pi^2 &=
  -\frac{1}{2M_B}
  \bra{B} \bar b_v (iD^\perp)^2 {b_v} \ket{B}, \notag \\
  \mu_G^2 &=
  \frac{1}{2M_B}
  \bra{B} \bar b_v (iD^\perp_\mu)(iD^\perp_\nu) (-i\sigma^{\mu\nu}) {b_v} \ket{B}, \notag \\
  \rho_{D}^3 &=
  \frac{1}{2M_B}
  \bra{B} \bar b_v (iD^\perp_\mu)
  (i v \cdot D)
  (iD^{\perp\nu}) {b_v} \ket{B}, \notag \\
  \rho_{LS}^3 &=
  \frac{1}{2M_B}
  \bra{B} \bar b_v (iD^\perp_\mu)
  (i v \cdot D)
  (iD^\perp_\nu) (-i\sigma^{\mu\nu}) {b_v} \ket{B},
\end{align}
where $D_\mu = \partial_\mu-ig_sA_\mu$, $D^\perp_\mu = (g_{\mu\nu} - v_\mu v_\nu)(iD^\nu)$, $b_v(x) = \exp(-i m_b v \cdot x) b(x)$.
The  $B$ meson velocity and mass are denoted by $v^\mu = p^\mu_B/m_B$ and $m_B$, respectively.

We consider two different approaches for the construction of the centralized moments:
\begin{enumerate}[label=(\Alph*)]
\item As a first step, expressions for centralized moments are obtained in the
  on-shell scheme.  To this end, the ratios in
  Eqs.~\eqref{eqn:CentralizedMomentsDefinition} and~\eqref{eq::MnDefinition_norm}
  are expanded up to $O(\alpha_s^3)$ (to leading order in $1/m_b$) and up to
  $1/m_b^3$ for the power corrections.  We discard higher $\alpha_s$
  corrections in the sub-leading power in $1/m_b$.  Afterwards one applies the
  transition to the kinetic scheme.

\item We convert the expressions for $Q_{i,j}$ and $L_i$ to the
  kinetic scheme.  In a second step the ratios in
  Eqs.~\eqref{eqn:CentralizedMomentsDefinition} and~\eqref{eq::MnDefinition_norm}
  are expanded up to $\alpha_s^3$ (to leading order in $1/m_b$) and up to
  $O(1/m_b^3)$ for the power corrections.
\end{enumerate}
Note that the two approaches do not yield the same analytic expressions
because of the redefinition of the HQE parameters, see
Eq.~(\ref{eqn:HQEParametersRedefinition}).  In approach (A) the perturbative
versions of $\mu_\pi$ and $\rho_D$ appear after expanding the centralized
moments in $\alpha_s$ and $1/m_b$.  In case (B) they are introduced before
expansion, and therefore treated as $\alpha_s$ corrections in the later
re-expansion of the ratios. Approach (A) and (B) start to differ at order
$\alpha_s^2$
since the shift of the power-suppressed terms according to
Eq.~(\ref{eqn:HQEParametersRedefinition})
induces perturbative $\alpha_s$ corrections from $1/m_b$ terms.
In both approaches, we retain all powers of the Wilsonian
cutoff $\mu/m_b^\kin$.  Only those terms involving one of the genuine
non-perturbative parameters are expanded in $1/m_b$.  For a further
discussion of the differences between the two approaches and their
interpretation we refer to Section~\ref{sec:AvsB} where ${\cal O}(\alpha_s)$
corrections to power-suppressed terms are considered for the $q^2$ moments.

We set the renormalization scale of the strong coupling constant
$\mu_s=m_b^\kin$ and use $\alpha_s^{(4)}(m_b^\kin)$ as expansion parameter,
i.e.\ we decouple the bottom quark from the running of $\alpha_s$, and we
re-expand in $\alpha_s^{(4)}$ up to third order.  We use the input values
\begin{align}
  m_b^\kin(1\, \GeV) &=4.526 \, \GeV, &
  \overline m_c (3\, \GeV) &=0.993 \, \GeV, \notag \\
  \mu & = 1 \, \GeV , &
  \alpha_s^{(4)}(m_b^\kin) &= 0.2186.
\end{align}
For the HQE parameters, we use the most updated values and their correlations
from~\cite{Bordone:2021oof}:
\begin{align}
  \mu_\pi^2  &=   0.477  (56) \, \GeV^2, &
  \rho_D^3 &= 0.185 (31)\, \GeV^3, \notag \\
  \mu_G^2 &= 0.306 (50) \, \GeV^2, &
  \rho_{LS}^3 &= -0.130 (92) \, \GeV^3,
  \label{eqn:HQEParameterValues}
\end{align}
where all parameters are defined at $\mu=1 \, \GeV$.

In the following we report the numerical prediction for the various moments in the kinetic scheme, 
employing approaches (A) and (B).
For each moment we factorize out the tree-level prediction,  
and show the size of the $\alpha_s, \alpha_s^2$ and $\alpha_s^3$ corrections 
(denoted by $X_{\alpha_s^n}$).
The quoted uncertainties come from the $\delta$ expansion as explained in the previous section.
We denote the sum of all $1/m_b^2$ and $1/m_b^3$ corrections by the subscript ``pw''.

For comparison, we quote also an uncertainty for the contribution of higher
$1/m_b$ corrections.  It arises from the uncertainties in the HQE parameters
given in Eq.~\eqref{eqn:HQEParameterValues} with correlations taken into
account.  We will use this uncertainty as reference value to compare the
relevance of the $\alpha_s^3$ corrections in the fits for $|V_{cb}|$.

\subsection{\boldmath $q^2$ moments}

We first show results for the $q^2$ moments with approach (A)
{\scalefont{0.9}
\begin{align} 
\hat q_1 &= &0.232947 &\Big[1-0.0106345_{\alpha_s} & & -0.008736(15)_{\alpha_s^2} & & -0.00505(13)_{\alpha_s^3} & & -0.0875(97)_{\mathrm{pw}}\Big], \notag \\ 
\hat q_2 &= &0.0235256 &\Big[1-0.035937_{\alpha_s} & & -0.0217035(20)_{\alpha_s^2} & & -0.01118(17)_{\alpha_s^3} & & -0.237(27)_{\mathrm{pw}}\Big], \notag \\ 
\hat q_3 &= &0.0014511 &\Big[1-0.0700381_{\alpha_s} & & -0.035693(73)_{\alpha_s^2} & & -0.01909(12)_{\alpha_s^3} & & -0.726(94)_{\mathrm{pw}}\Big], \notag \\ 
\hat q_4 &= &0.00120161 &\Big[1-0.0585199_{\alpha_s} & & -0.042276(11)_{\alpha_s^2} & & -0.02411(20)_{\alpha_s^3} & & -0.631(77)_{\mathrm{pw}}\Big]. 
\end{align} 
}
With approach (B) we obtain:
{\scalefont{0.9}
\begin{align} 
\hat q_1 &= &0.232947 &\Big[1-0.0106332_{\alpha_s} & & -0.007100(16)_{\alpha_s^2} & & -0.00326(13)_{\alpha_s^3} & & -0.0875(97)_{\mathrm{pw}}\Big], \notag \\ 
\hat q_2 &= &0.0235256 &\Big[1-0.0359328_{\alpha_s} & & -0.0175591(28)_{\alpha_s^2} & & -0.00677(17)_{\alpha_s^3} & & -0.237(27)_{\mathrm{pw}}\Big], \notag \\ 
\hat q_3 &= &0.00145109 &\Big[1-0.0700256_{\alpha_s} & & -0.030529(71)_{\alpha_s^2} & & -0.01282(12)_{\alpha_s^3} & & -0.726(94)_{\mathrm{pw}}\Big], \notag \\ 
\hat q_4 &= &0.0012016 &\Big[1-0.0585099_{\alpha_s} & & -0.0342994(88)_{\alpha_s^2} & & -0.01597(20)_{\alpha_s^3} & & -0.631(77)_{\mathrm{pw}}\Big]. 
\end{align}

}

For the $q^2$ moments we observe a good behaviour of the perturbative series,
with coefficients precisely determined via the $\delta$-expansion.
Note that for the $q^2$ moments, even $\alpha_s^2$ corrections are not  yet
available in the literature as the results presented in Refs.~\cite{Biswas:2009rb,Gambino:2011cq} are only for
electron energy and hadronic invariant mass moments.

The size of the $\alpha_s^2$ corrections are of few percent while
third order corrections are about a factor of two smaller and in 
the range of $0.5- 2\%$. We observe that higher power corrections
are sizable and as large as 70\% of the leading order contribution.
The estimated uncertainty of the power corrections are a factor two 
to three larger compared to the $\alpha_s^3$ term. 
At $O(\alpha_s^3)$ the difference between the two approaches yields a difference of
$0.3\%,0.9\%,1.1\%$ and $1.6\%$ for the four moments which is of the same order of magnitude as the
$\alpha_s^3$ terms.

Central moments of the $q^2$ spectrum have been measured recently by
Belle~\cite{Belle:2021idw} separately for electrons and muons in the final
state.  The quoted results for a cut on the leptonic invariant mass of
$q^2> 3 \, \GeV^2$, averaged between muon and electron, read\footnote{We thank
  F.\ Bernlochner and R.\ van Tonder for providing us with the values of the
  centralized moments constructed from the data of Ref.~\cite{Belle:2021idw}.}
\begin{align}
  q_1 (q^2>3 \, \GeV^2) &= 6.23 \, (8) \, \GeV^2, \notag \\
  q_2 (q^2>3 \, \GeV^2) &= 4.44 \, (15) \, \GeV^4, \notag \\
  q_3 (q^2>3 \, \GeV^2) &= 4.13 \, (68) \, \GeV^6, \notag \\
  q_4 (q^2>3 \, \GeV^2) &= 46.6 \, (5.6) \, \GeV^8\,.
                          \label{eq::q2_Belle}
\end{align}
Due to the cut of $q^2$ we refrain from a direct comparison
to our predictions. However, it is interesting to compare the uncertainties.
The moments in Eq.~(\ref{eq::q2_Belle})
have a relative uncertainty of $1.3\%, 3.1\%, 16\%$ and $12\%$.  The
experimental error of $q_1$ and $q_2$ is only about a factor two larger
compared to the magnitude of the $\alpha_s^3$ term. Furthermore, note
that the measurements in~\cite{Belle:2021idw} with a higher cut on $q^2$ have
even smaller uncertainties reaching a precision of $0.5\%$ which
makes the $\alpha_s^3$ corrections even more relevant.

\subsection{Charged-Lepton Energy Moments}
For the electron energy moments our result in the approach (A) read
{\scalefont{0.9}
\begin{align} 
\hat \ell_1 &= &0.315615 &\Big[1-0.0101064_{\alpha_s} & & -0.005082(17)_{\alpha_s^2} & & -0.00227(13)_{\alpha_s^3} & & -0.0192(31)_{\mathrm{pw}}\Big], \notag \\ 
\hat \ell_2 &= &0.00900585 &\Big[1-0.01992_{\alpha_s} & & -0.006152(41)_{\alpha_s^2} & & +0.0002(21) _{\alpha_s^3} & & +0.017(11) _{\mathrm{pw}}\Big], \notag \\ 
\hat \ell_3 &= &-0.000464269 &\Big[1-0.0639319_{\alpha_s} & & -0.035673(10)_{\alpha_s^2} & & -0.0142(46)_{\alpha_s^3} & & -0.175(22)_{\mathrm{pw}}\Big], \notag \\ 
\hat \ell_4 &= &0.00020743 &\Big[1-0.028854_{\alpha_s} & &
                                                           -0.00717(23)_{\alpha_s^2}
                                                                                     & & -{0.00(25)} _{\alpha_s^3} & & +{0.000(21)} _{\mathrm{pw}}\Big]\,,
\label{eq::lepmom_A}
\end{align} 
}
while for (B) we find
{\scalefont{0.9}
\begin{align} 
\hat \ell_1 &=\!\!\! &0.315615 &\Big[1-0.010106_{\alpha_s} & & -0.004838(17)_{\alpha_s^2} & & -0.00200(13)_{\alpha_s^3} & & -0.0192(31)_{\mathrm{pw}}\Big], \notag \\ 
\hat \ell_2 &=\!\!\! &0.00900585 &\Big[1-0.0199202_{\alpha_s} & & -0.006303(42)_{\alpha_s^2} & & -0.0001(21) _{\alpha_s^3} & & +0.017(11) _{\mathrm{pw}}\Big], \notag \\ 
\hat \ell_3 &=\!\!\! &-0.000464268 &\Big[1-0.0639261_{\alpha_s} & & -0.0358480(91)_{\alpha_s^2} & & -0.0142(46)_{\alpha_s^3} & & -0.175(22)_{\mathrm{pw}}\Big], \notag \\ 
\hat \ell_4 &=\!\!\! &0.00020743 &\Big[1-0.0288534_{\alpha_s} & &
                                                            -0.00611(23)_{\alpha_s^2}
                                                                                    &
                                                                                      & +0.00(25) _{\alpha_s^3} & & +0.000(21) _{\mathrm{pw}}\Big]. 
\label{eq::lepmom_B}
\end{align} 
}

For these moments we observe in general a good convergence of the perturbative
series in the kinetic scheme. It is interesting to note that the relative size
of the $\alpha_s^3$ corrections are smaller compared to those found for $q^2$
moments.  For $\hat \ell_1$ and $\hat \ell_2$ we have 0.2\% and 0.02\% and for
$\hat \ell_3$ about 1.4\%.  For $\hat \ell_4$, the $\alpha_s^3$ correction is
not determined in a reliable way due to the uncertainty of the finite
expansion in $\delta$.  On the other hand, also the impact of the power
corrections is much smaller compared to $q^2$ moments.  For $\hat \ell_1$ and
$\hat \ell_2$ the power correction uncertainty is of the order of 0.1-0.3\%
and comparable with the size of $\alpha_s^3$ corrections.  The $\alpha_s^3$
coefficient of $\hat \ell_2$ is small which is likely due to numerical
cancellation.  In case of $\hat \ell_3$ the uncertainty coming from higher
$1/m_b$ terms of about $2.2\%$ is comparable with the $\alpha_s^3$ correction.

The difference between our predictions obtained with the approaches (A) and (B) are small, 
and overall they never exceed the $0.1\%$ of the leading order contribution. 

We can examine the precision of experimental measurements for instance by
quoting the values of the electron energy moments, with a cut $E_\ell>0.4$~GeV, 
as measure by Belle~\cite{Belle:2006kgy}
\begin{align}
  \ell_1(E_\ell > 0.4 \, \GeV) &=  1393.92  ( 6.73 )( 3.02) \, \mathrm{MeV}, \notag \\
  \ell_2(E_\ell > 0.4 \, \GeV) &=  168.77  ( 3.68 )( 1.53 ) \times 10^{-3}\, \GeV^2,\notag \\
  \ell_3(E_\ell > 0.4 \, \GeV) &=  -21.04  ( 1.93 ) ( 0.66 ) \times 10^{-3}\, \GeV^3, \notag \\
  \ell_4(E_\ell > 0.4 \, \GeV) &=  64.153  ( 1.813 )( 0.935 ) \times 10^{-3}\, \GeV^4. 
                                 \label{eq::mom_el_belle}
\end{align}
The relative accuracies of these measurements are $0.5\%, 2.3\%, 9.6\%$ and
$3.2\%$, respectively.  Due to the applied cut, the central values cannot
directly be compared to our prediction.  However, we note that for
$\hat \ell_1$ the $\alpha_s^3$ corrections are only a factor of two smaller
than the experimental error.  Also for the moments of the charged lepton
energy, the experimental measurements are in general more precise at higher
values of the cut.  Therefore for some of the moments, third order QCD
corrections are already comparable to the experimental error and the
uncertainties associated to power corrections.

\subsection{Hadronic Invariant Mass Moments}

Finally let us analyze the predictions for the hadronic invariant mass
moments.  For approach (A) we have 
{\scalefont{0.9}
\begin{align} 
\hat h_1 &= &0.00899843 &\Big[+23.4975 & & + 1 +0.4223(15)_{\alpha_s^2} & & +0.147(11)_{\alpha_s^3} & & +0.04(20) _{\mathrm{pw}}\Big], \notag \\ 
\hat h_2 &= &0.000745468 &\Big[+0.87352 & & + 1 +0.4505(74)_{\alpha_s^2} & & +0.34(43) _{\alpha_s^3} & & +3.33(59)_{\mathrm{pw}}\Big], \notag \\ 
\hat h_3 &= &0.0000915954 &\Big[-0.0729568 & & + 1 +0.165(62)_{\alpha_s^2} & & +2.29(55)_{\alpha_s^3} & & +7.3(1.1) _{\mathrm{pw}}\Big], \notag \\ 
\hat h_4 &= &0.000091207 &\Big[+0.0100938 & & + 1 +0.51(17) _{\alpha_s^2} & & +1 (145) _{\alpha_s^3} & & +0.380(52)_{\mathrm{pw}}\Big]\,,
\end{align} 
} 
while for (B) we find 
{\scalefont{0.9}
\begin{align} 
\hat h_1 &= &0.00899836 &\Big[+23.4976 & & + 1 +0.4114(15)_{\alpha_s^2} & & +0.134(11)_{\alpha_s^3} & & +0.04(20) _{\mathrm{pw}}\Big], \notag \\ 
\hat h_2 &= &0.000745462 &\Big[+0.873533 & & + 1 +0.3971(73)_{\alpha_s^2} & & +0.25(43) _{\alpha_s^3} & & +3.33(59)_{\mathrm{pw}}\Big], \notag \\ 
\hat h_3 &= &0.0000915935 &\Big[-0.0729428 & & + 1 -0.088(61) _{\alpha_s^2} & & +2.00(55)_{\alpha_s^3} & & +7.3(1.1) _{\mathrm{pw}}\Big], \notag \\ 
\hat h_4 &= &0.0000912064 &\Big[+0.0100992 & & + 1 +0.56(16) _{\alpha_s^2} & & +0(145) _{\alpha_s^3} & & +0.380(52)_{\mathrm{pw}}\Big]. 
\label{eq::hadr_mom_B}
\end{align} 
}
As before, we
normalize the various higher order terms w.r.t.\ the $O(\alpha_s)$
corrections, since the partonic tree-level invariant mass vanishes.

Our approximation does not determine $\hat h_4$ at $O(\alpha_s^3)$ and also
for $\hat h_2$ we can only provide the order of magnitude.  While for
$\hat h_1$ and $\hat h_2$ the perturbative series still displays a good
convergence, the prediction for $\hat h_3$ shows an enhanced $O(\alpha_s^3)$
term which is more than a factor of two larger than the $O(\alpha_s)$ 
contribution.  For $\hat h_3$ also the power-suppressed terms are quite
large and the corresponding uncertainty is as large as the $O(\alpha_s)$
term.  This calls for a careful assessment of the theoretical uncertainties
for this specific moment, or as a conservative approach, for the elimination
of $\hat h_3$ from the set of observables considered in the fits.  For
$\hat h_1$ the relative difference between approaches (A) and (B) is about
$0.1\%$ while for $\hat h_2$ and $\hat h_3$ it is of $2.3\%$ and $5\%$,
respectively.

From the expressions in Eq.~(\ref{eq::hadr_mom_B}) we obtain after multiplication
with the proper power of $m_b$ the results
\begin{align}
h_1 &= 4.63 (4) \GeV ^2, \notag \\
h_2 &= {1.88} (23) \GeV^4 , \notag \\
h_3 &= {8.41 (97)} \GeV^6,
\label{eq::h123}
\end{align}
where the uncertainties in Eq.~(\ref{eq::hadr_mom_B}) have been added in
quadrature. We refrain from listing $h_4$ since there is a strong dependence
on the higher order power-suppressed corrections~\cite{Gambino:2016jkc}.
The results in Eq.~(\ref{eq::h123}) can be compared to the
experimental measurements of the $M_X$ moments performed by
DELPHI~\cite{DELPHI:2005mot}:\footnote{We thank P.\ Gambino for
  clarification about the value of $h_1$.}
\begin{align}
  h_1 &= 4.541\, (101)  \, \GeV^2, \notag \\
  h_2 &= 1.56 \, (18)\,  (16) \, \GeV^4,\notag \\
  h_3 &= 4.05\,  (74)\,  (32) \, \GeV^6. 
         \label{eq::h123_exp}
\end{align}
Note that no cuts have been applied.  Their relative errors are $2\%$,$15\%$
and $20\%$, respectively.  For $h_1$ one observes agreement within the
uncertainties. Note, however, that the experimental error is about a factor
2.5 larger than the one from the theory prediction. Furthermore, from
Eq.~(\ref{eq::hadr_mom_B}) one observes that the contribution from the
$\alpha_s^3$ term has about the same order of magnitude as the theory
uncertainty.  Also for $h_2$ we find agreement between the theory prediction
and the experimental result. However, one has to keep in mind that the theory
prediction is dominated by the power-suppressed terms.  In the case of $h_3$
it is worth mentioning that the expansion in $\alpha_s$ does not
converge. Furthermore, there are large contributions from the power-suppressed
terms and thus it is not surprising that the numbers in Eqs.~(\ref{eq::h123})
and~(\ref{eq::h123_exp}) do not agree within the one sigma range of the
uncertainties.  Let us mention that for $h_2$ and $h_3$ we observe that the
$\alpha_s^3$ terms are larger than the quoted error by DELPHI.



\section{\boldmath Including NLO perturbative corrections to the\\ power suppressed terms}
\label{sec:AvsB}

In this section we study the origin in the numerical differences between
approach (A) and (B), and how it can be reduced by including NLO perturbative
corrections to the power suppressed terms, i.e.\ by taking into account
$O(\alpha_s)$ corrections in the Wilson coefficients of the HQE parameters
$\mu_\pi^2,\mu_G^2,\rho_D^3$ and $\rho_{LS}^3$.  We will refer to these
correction as $\alpha_s/m_b^n$ corrections ($n=2$ or $3$ in our case).

We focus on the $q^2$ moments.  Analytic results for the $q^2$ spectrum
including $\alpha_s/m_b^n$ corrections were recently computed
in~\cite{Mannel:2021zzr}.  By performing an analytic integration of the
differential decay rate, we obtain expressions for the perturbative
corrections to power suppressed terms of the $q^2$ moments.  Schematically
they have the form (compare also with
Eq.~\eqref{eqn:PerturbativeExpansionMoments})
\begin{align}
  Q_{i,0} &=
   \Bigg[
    Q_{i,0}^{(0)}
  + Q_{i,0}^{(1)} \frac{\alpha_s}{\pi}
  + Q_{i,0}^{(2)} \left(\frac{\alpha_s}{\pi}\right)^2
  + Q_{i,0}^{(3)} \left(\frac{\alpha_s}{\pi}\right)^3
  \Bigg]
  \left(1-\frac{\mu_\pi^2}{2m_b^2} \right)
  + Q_{i,0, \mu_G}^{(0)} \left( \frac{\mu_G^2}{m_b^2}-\frac{\rho_{LS}^3}{m_b^3} \right)
  \notag \\[5pt] & \quad
  + Q_{i,0, \mu_G}^{(1)} \frac{\alpha_s}{\pi} \frac{\mu_G^2}{m_b^2}
  + Q_{i,0, \rho_{LS}}^{(1)} \frac{\alpha_s}{\pi}  \frac{\rho_{LS}^3}{m_b^3}
  +\Bigg[
     Q_{i,0, \rho_D}^{(0)}
    +Q_{i,0, \rho_D}^{(1)} \frac{\alpha_s}{\pi}
  \Bigg]
  \frac{\rho_D^3}{m_b^3}\,,
  \label{eqn:apipw}
\end{align}
where $\alpha_s\equiv \alpha_s(\mu_s)$. For convenience we provide analytic
results for $Q_{i,0, \mu_G}^{(0)}$, $Q_{i,0, \mu_G}^{(1)}$ and
$Q_{i,0, \rho_{LS}}^{(1)}$ in Appendix~\ref{app::Qi0}.  The results for
$Q_{i,0, \rho_D}^{(0)}$ and $Q_{i,0, \rho_D}^{(1)}$ can be found in
Ref.~\cite{Mannel:2021zzr}.

Let us compare the predictions for the centralized moments $q_i$ obtained 
in Eqs. (33) and (34) where no $O(\alpha_s/m_b^n)$ correction was taken into account. We obtain
\begin{align}
  \Delta q_1 &= {0.3}\%,\notag \\
  \Delta q_2 &= {0.9}\%, \notag \\
  \Delta q_3 &= {1.1}\%,\notag \\
  \Delta q_4 &= {1.6}\%,
               \label{eq::diff_1}
\end{align}
where we define the relative difference between scheme (A) and (B) by
\begin{equation}
  \Delta q_i \equiv  \frac{{|\hat{q}_i^{(A)}-\hat{q}_i^{(B)}|}}{\hat{q}_i^{\mathrm{LO}}}.
  \label{eqn:Deltaqi}
\end{equation}
Let us explain the origin of such difference.
It is related to terms of the form
\begin{equation}
  \alpha_s \times \frac{\rho_D^3}{m_b^3}.
\end{equation}
In the kinetic scheme one has to redefine $\rho_D^3(0)$
according to Eq.~(\ref{eqn:HQEParametersRedefinition})
where the perturbative expansion of
$[\rho_D(\mu)]_{\mathrm{pert}}$ is given by
\begin{equation}
  [\rho_D^3(\mu)]_{\mathrm{pert}} =
  \mu^3
  \sum_{n\ge 1}
  r_{\mathrm{pert}}^{(n)} \left( \frac{\alpha_s}{\pi} \right)^n.
\end{equation}
The coefficients $r_{\mathrm{pert}}^{(n)}$ are known up to $O(\alpha_s^3)$
from~\cite{Fael:2020njb}. Their explicit expressions are not relevant for our
discussion.  For $q^2$ moments we can ignore the role of $\mu_\pi^2$ since its
dependence drops out due to reparametrization invariance~\cite{Fael:2018vsp}.

In case we neglect terms of $O(\alpha_s /m_b^n)$, contributions scaling like
$\alpha_s \times \rho_D^3/m_b^3$ are dropped in approach~(A) after
re-expansion of~\eqref{eqn:CentralizedMomentsDefinition} in the on-shell
scheme. In approach~(B) we first transform the building blocks entering
Eq.~\eqref{eqn:CentralizedMomentsDefinition} to the kinetic scheme. In
particular, we redefine $\rho_D^3$ according to
Eq.~(\ref{eqn:HQEParametersRedefinition}).  After inserting the expressions in
Eq.~\eqref{eqn:CentralizedMomentsDefinition} and expanding in $\alpha_s$ new
terms of order $\alpha_s^2$ are generated since the ratio $\mu^3/m_b^3$ is
considered of order one and not $1/m_b^3$. Thus, we observe that the
difference between (A) and (B) scales like
$\alpha_s^2 r_{\mathrm{pert}}^{(1)} \mu^3/m_b^3$ if $O(\alpha_s /m_b^n)$ terms
are neglected. In case $\alpha_s/m_b^n$ terms are included the difference is
of order $\alpha_s^2 r_{\mathrm{pert}}^{(1)} \mu^6/m_b^6$.

We now compare the values of the $q^2$ moments obtained in approaches (A) 
and (B) after the inclusion of terms of $O(\alpha_s/m_b^n)$.
We recompute the prediction for the centralized moments $q_i$
by re-expanding the final result up to $O(\alpha_s^3)$ at the partonic level, 
while we keep corrections of $O(\alpha_s/m_b^n)$ in the power 
suppressed terms.
With approach (A) we obtain
{\scalefont{0.9}
\begin{align} 
\hat q_1 &= &0.232947 &\Big[1-0.0106137_{\alpha_s} & & -0.00383463_{\alpha_s^2} & & -0.00327(13)_{\alpha_s^3} & & -0.097(11)_{\mathrm{pw}}\Big], \notag \\ 
\hat q_2 &= &0.0235256 &\Big[1-0.0359242_{\alpha_s} & & -0.00697531_{\alpha_s^2} & & -0.00683(17)_{\alpha_s^3} & & -0.240(27)_{\mathrm{pw}}\Big], \notag \\ 
\hat q_3 &= &0.0014511 &\Big[1-0.0701143_{\alpha_s} & & +0.0145548_{\alpha_s^2} & & -0.00866(13)_{\alpha_s^3} & & -0.624(80)_{\mathrm{pw}}\Big], \notag \\ 
\hat q_4 &= &0.00120161 &\Big[1-0.058515_{\alpha_s} & & -0.000100666_{\alpha_s^2} & & -0.01686(20)_{\alpha_s^3} & & -0.545(65)_{\mathrm{pw}}\Big]\,,
\end{align} 
}
and approach (B) leads to
{\scalefont{0.9}
\begin{align} 
\hat q_1 &= &0.232947 &\Big[1-0.0106265_{\alpha_s} & & -0.00402646_{\alpha_s^2} & & -0.00190(13)_{\alpha_s^3} & & -0.094(11)_{\mathrm{pw}}\Big], \notag \\ 
\hat q_2 &= &0.0235256 &\Big[1-0.0359104_{\alpha_s} & &
                                                        -0.00817945_{\alpha_s^2} & & -0.00366(17)_{\alpha_s^3} & & -0.227(26)_{\mathrm{pw}}\Big], \notag \\ 
\hat q_3 &= &0.00145109 &\Big[1-0.0699819_{\alpha_s} & &
                                                         +0.00342844_{\alpha_s^2} & & -0.00822(12)_{\alpha_s^3} & & -0.510(68)_{\mathrm{pw}}\Big], \notag \\ 
\hat q_4 &= &0.0012016 &\Big[1-0.0584734_{\alpha_s} & &
                                                        -0.00681918_{\alpha_s^2} & & -0.01185(20)_{\alpha_s^3} & & -0.477(58)_{\mathrm{pw}}\Big]. 
\end{align} 
}
Taking the difference from leading $m_b$ contribution only, i.e., from the
terms flagged by ``$\alpha_s^i$'' we obtain
\begin{align}
  \Delta q_1 &= {0.1}\%,\notag \\
  \Delta q_2 &= {0.2}\%, \notag \\
  \Delta q_3 &= {1.1}\%,\notag \\
  \Delta q_4 &= { 0.2}\%.
               \label{eq::diff_2}
\end{align}
Comparing Eqs.~(\ref{eq::diff_2}) and~(\ref{eq::diff_1}) we observe that the
predictions using (A) and (B) get closer after the inclusion of the 
$O(\alpha_s/m_b^n)$ corrections.  This happens because now both approach (A)
and (B) take into account contributions scaling as
$\alpha_s \times \rho_D^3/m_b^3$.  After redefinition of $\rho_D$, both (A)
and (B) generate the same corrections of the form
$\alpha_s^2 r^{(1)}_{\mathrm{pert}} \mu^3/m_b^3$.  Therefore $\Delta q_i$
become smaller.

However if we take into account also the power-suppressed terms,
i.e., the parts flagged by ``pw'', we obtain
\begin{align}
  \Delta q_1 &= {0.4}\%,\notag \\
  \Delta q_2 &= {1.5}\%, \notag \\
  \Delta q_3 &= {10.3}\%,\notag \\
  \Delta q_4 &= {6.6}\%,
\end{align}
which are even larger than without including $O(\alpha_s/m_b^n)$ terms.
Similarly to what we observed before, the difference starts now at order
$1/m_b^2$ and $1/m_b^3$ because of contributions of the form
\begin{equation}
  \alpha_s \times \frac{[\rho_D^3]_{\mathrm{pert}}}{m_b^3} \times
  \frac{\mu_G^2}{m_b^2}
  \quad 
  \text{or}
  \quad 
  \alpha_s \times \frac{[\rho_D^3]_{\mathrm{pert}}}{m_b^3} \times
  \frac{\rho_D^3(\mu)}{m_b^3}
\end{equation}
which arise if one uses approach (B).
However these terms are actually of $O(1/m_b^5)$ and $O(1/m_b^6)$ and therefore
they would not appear if $[\rho_D]_{\mathrm{pert}}/m_b^3 \sim \mu^3/m_b^3$ 
is considered as a $1/m_b^3$ suppressed term and the expressions for the moments
re-expanded up to $1/m_b^3$.

In the end, we conclude that the ambiguity between approaches (A) and (B) can
be removed if the power corrections $\mu/m_b$ originating from the kinetic
scheme are considered as $1/m_b$ suppressed term in the HQE.  Note that for
the charged-lepton energy moments the contribution from the power-suppressed
terms are significantly smaller and thus the different treatment of the
$\mu/m_b$ terms is numerically less important as can be seen from the
comparison of Eqs.~(\ref{eq::lepmom_A}) and~(\ref{eq::lepmom_B}).




\section{Conclusions}
\label{sec:conclusions}

In this work we compute several kinematic moments of inclusive
$B \to X_c \ell \bar \nu_\ell$ decays up to $O(\alpha_s^3)$.  In particular we
consider for the first time higher order QCD corrections to $q^2$ moments.  We
use the optical theorem to obtain analytic expressions for the moments as an
expansion in the parameter $\delta=1-m_c/m_b$.  For most of the considered
observables, the series expansion in $\delta$ is sufficient to obtain precise
results for the coefficients of the perturbative expansion.  However, for some
of the centralized moments, there are significant cancellations and our finite
expansion depth in $\delta$ does not allow for a determination of the
$\alpha_s^3$ corrections in a reliable way.
Note that also a calculation based on numerical methods might have
similar problems since also there in a first step the elementary moments are
computed with a finite numerical accuracy~\cite{Gambino:2011cq}.

We describe in detail our computational methods. The quark
masses are renormalized in the on-shell scheme.  Afterwards, we study the
moments in the kinetic scheme and investigate the importance of the higher
order QCD corrections for the determination of $|V_{cb}|$. To this end, we
present numerical results in the kinetic scheme together with the contribution
from higher $1/m_b$ power corrections and the related uncertainties.

For the first two $q^2$ and electron energy moments, we find that the third
order corrections are of the same order as the uncertainties associated to
$1/m_b^2$ and $1/m_b^3$ corrections. Furthermore, they are comparable in size
with experimental errors.  Thus, the inclusion of $\alpha_s^3$ corrections in
future analyses might be important.  For the hadronic invariant mass moments
$\hat h_2$ and $\hat h_3$ we observe $\alpha_s^3$ corrections which are of the
same order of magnitude or even larger than experimental uncertainties and
thus might influence the $|V_{cb}|$ fit. For these moments also the
power-suppressed terms are sizeable.

We discuss two approaches for the construction of the centralized moments in
the kinetic scheme. In approach (A) the scheme transformation rules are
applied to the centralized moments in the on-shell scheme. On the other hand,
in approach (B) the building blocks are transformed to the kinetic scheme and
the centralized moments are constructed afterwards.  The numerical results
differ starting from order $\alpha_s^2$ which is due to the fact that
$\mu/m_b$ counts as order one, where $\mu$ is the Wilsonian cutoff of the
kinetic scheme.  For the $q^2$ we show that the difference reduces in case
higher order QCD corrections to the power-suppressed terms are considered.

The analysis of the inclusive third order corrections of charged-lepton
energy, leptonic invariant mass and hadronic invariant mass moments performed
in this paper suggests that one should initiate a differential calculations at
third order.


\section*{Acknowledgements}  

We kindly thank Alexander Smirnov for providing us with the development
version of {\tt FIRE}. We also thank Joshua Davies for many useful hints on
the efficient treatment of large expressions with {\tt FORM}. Furthermore, we
are grateful to Paolo Gambino for useful comments to the manuscript.  Feynman
diagrams were drawn with the help of Axodraw~\cite{Vermaseren:1994je} and
JaxoDraw~\cite{Binosi:2003yf}.  This research was supported by the Deutsche
Forschungsgemeinschaft (DFG, German Research Foundation) under grant 396021762
--- TRR 257 ``Particle Physics Phenomenology after the Higgs Discovery''.

\appendix


\section{\label{app::tensor}Tensor decomposition formulas}

In this appendix we report the formulas employed to compute one-loop hard and
ultra-soft tensor integrals.  We denote by
$\{[g]^r [p]^{N-2r}\}^{\mu_1 \dots \mu_N}$ the product of $r$ metric tensors
and $N-2r$ vectors $p$, totally symmetric in its $N$ Lorentz indices.

\subsection{Massless two-point integral}
The tensor integral of a massless one-loop two-point function is given by (see e.g.\ Ref.~\cite{Smirnov:2012gma})
\begin{multline}
  \int \frac{d^d k}{(2\pi)^d}
  \frac{k^{\mu_1}\dots k^{\mu_N}}{(-k^2)^{n_1}(-(k-q))^{n_2}} =
  \frac{i}{(4\pi)^{d/2}}(-q^2)^{d/2-n_1-n_2}  \\[5pt]
  \sum_{r=0}^{[N/2]} 
  \frac{\Gamma(n_1+n_2-r-d/2)\Gamma(d/2+N-n_1-r) \Gamma(d/2-n_2+r)}
  {2^r \Gamma(n_1) \Gamma(n_2)\Gamma(d+N-n_1-n_2)}
  (q^2)^{r} \{[g]^r [q]^{N-2r}\}^{\mu_1 \dots  \mu_N} ,
  \label{eqn:tdecmassless}
\end{multline}
where $[N/2]$ is the greatest integer less than or equal to $N/2$.

\subsection{On-shell two-point integral with one mass}
The tensor integral of a massive one-loop two-point function  reads
\begin{multline}
  \int \frac{d^d k}{(2\pi)^d}
  \frac{k^{\mu_1}\dots k^{\mu_N}}{(-k^2)^{n_1}(-k^2+2 p\cdot k)^{n_2}} =
  \frac{i}{(4\pi)^{d/2}}(m^2)^{d/2-n_1-n_2}  \\[5pt]
  \sum_{r=0}^{[N/2]} 
  \frac{\Gamma(n_1+n_2-r-d/2)\Gamma(d+N-2n_1-n_2)}
  {(-2)^r \Gamma(n_2)\Gamma(d+N-n_1-n_2)}
  (m^2)^{r} \{[g]^r [p]^{N-2r}\}^{\mu_1 \dots  \mu_N} 
  \label{eqn:tdecmassive}
\end{multline}
where $p^2 = m^2$.
Such integrals appear in case the loop momentum is hard.

\subsection{Ultra-soft integral}
The tensor integral of a one-loop ultra-soft two-point function is given by
\begin{multline}
  \int \frac{d^d k}{(2\pi)^d}
  \frac{k^{\mu_1}\dots k^{\mu_N}}{(-k^2)^{n_1}(-2 p\cdot k+y)^{n_2}} =
  \frac{i}{(4\pi)^{d/2}}y^{d-2n_1-n_2+N} (p^2)^{n_1-N-d/2}  \\[5pt]
  \sum_{r=0}^{[N/2]} 
  \frac{(-1)^{N+r}
  \Gamma(d/2-n_1-r+N)\Gamma(2n_1+n_2-N-d)}{2^r \Gamma(n_1)\Gamma(n_2)}
  (p^2)^{r} \{[g]^r [p]^{N-2r}\}^{\mu_1 \dots  \mu_N} .
  \label{eqn:tdecultrasoft}
\end{multline}


\section{\label{app::Qi0}Inclusive $q^2$ moments to order $\alpha_s$}

The analytic results for the leading $m_b$ expansion terms read
\begin{align}
  Q_{0,0}^{(1)} &=
  C_F \Biggl\{
        \frac{25}{8}
        -\frac{239 \tilde{\rho} }{6}
        +\frac{239 \tilde{\rho} ^3}{6}
        -\frac{25 \tilde{\rho} ^4}{8}
        +\ln (\tilde{\rho} ) \biggl(
                -10 \tilde{\rho} 
                -45 \tilde{\rho} ^2
                +\frac{2 \tilde{\rho} ^3}{3}
                -\frac{17 \tilde{\rho} ^4}{6}
                \nonumber \\ &
                +64 \tilde{\rho} ^{3/2} (1+\tilde{\rho} ) \ln \big(1+\sqrt{\tilde{\rho} }\big)
        \biggr)
        +\ln (1-\tilde{\rho} ) \biggl[
                -\frac{17}{6}
                +\frac{32 \tilde{\rho} }{3}
                -\frac{32 \tilde{\rho} ^3}{3}
                +\frac{17 \tilde{\rho} ^4}{6}
                \nonumber \\ &
                +\biggl(
                        2+60 \tilde{\rho} ^2+2 \tilde{\rho} ^4-32 \tilde{\rho} ^{3/2}-32 \tilde{\rho} ^{5/2}\biggr) \ln (\tilde{\rho} )
        \biggr]
        +\pi ^2 \biggl(
                -\frac{1}{2}-8 \tilde{\rho} ^2-\frac{\tilde{\rho} ^4}{2}+16 \tilde{\rho} ^{3/2}
                \nonumber \\ &
                +16 \tilde{\rho} ^{5/2}\biggr)
        -\frac{1}{2} \tilde{\rho} ^2 \big(
                36+\tilde{\rho} ^2\big) \ln ^2(\tilde{\rho} )
        -128 \tilde{\rho} ^{3/2} (1+\tilde{\rho} ) \text{Li}_2\big(
                \sqrt{\tilde{\rho} }\big)
        +\big(
                3+48 \tilde{\rho} ^2+3 \tilde{\rho} ^4
                \nonumber \\ &
                +32 \tilde{\rho} ^{3/2}+32 \tilde{\rho} ^{5/2}\big) \text{Li}_2(\tilde{\rho} )
  \Biggr\}\,,
  \\
  Q_{1,0}^{(1)} &=
  C_F \Biggl\{
        \frac{39}{40}
        -\frac{17117 \tilde{\rho} }{600}
        -\frac{139129 \tilde{\rho} ^2}{900}
        +\frac{139129 \tilde{\rho} ^3}{900}
        +\frac{17117 \tilde{\rho} ^4}{600}
        -\frac{39 \tilde{\rho} ^5}{40}
        \nonumber \\ &
        +\ln (1-\tilde{\rho} ) \biggl[
                -\frac{301}{300}
                +\frac{185 \tilde{\rho} }{36}
                +\frac{50 \tilde{\rho} ^2}{3}
                -\frac{50 \tilde{\rho} ^3}{3}
                -\frac{185 \tilde{\rho} ^4}{36}
                +\frac{301 \tilde{\rho} ^5}{300}
                +\ln (\tilde{\rho} ) \biggl(
                        \frac{3}{5}+\frac{\tilde{\rho} }{3}
                        \nonumber \\ &
                        +110 \tilde{\rho} ^2+110 \tilde{\rho} ^3+\frac{\tilde{\rho} ^4}{3}+\frac{3 \tilde{\rho} ^5}{5}-32 \tilde{\rho} ^{3/2}-\frac{2368}{15} \tilde{\rho} ^{5/2}-32 \tilde{\rho} ^{7/2}\biggr)
        \biggr]
        +\ln (\tilde{\rho} ) \biggl[
                -\frac{123 \tilde{\rho} }{20}
                \nonumber \\ &
                -\frac{3251 \tilde{\rho} ^2}{30}
                -\frac{917 \tilde{\rho} ^3}{10}
                -\frac{91 \tilde{\rho} ^4}{90}
                -\frac{301 \tilde{\rho} ^5}{300}
                +\biggl(
                        64 \tilde{\rho} ^{3/2}+\frac{4736}{15} \tilde{\rho} ^{5/2}+64 \tilde{\rho} ^{7/2}\biggr) \ln \big(1+\sqrt{\tilde{\rho} }\big)
        \biggr]
        \nonumber \\ &
        +\pi ^2 \biggl(
                -\frac{3}{20}-\frac{\tilde{\rho} }{12}-16 \tilde{\rho} ^2-16 \tilde{\rho} ^3-\frac{\tilde{\rho} ^4}{12}-\frac{3 \tilde{\rho} ^5}{20}+16 \tilde{\rho} ^{3/2}+\frac{1184}{15} \tilde{\rho} ^{5/2}+16 \tilde{\rho} ^{7/2}\biggr)
        \nonumber \\ &
        +\text{Li}_2(\tilde{\rho} ) \biggl(
                \frac{9}{10}+\frac{\tilde{\rho} }{2}+96 \tilde{\rho} ^2+96 \tilde{\rho} ^3+\frac{\tilde{\rho} ^4}{2}+\frac{9 \tilde{\rho} ^5}{10}+32 \tilde{\rho} ^{3/2}
                +\frac{2368}{15} \tilde{\rho} ^{5/2}+32 \tilde{\rho} ^{7/2}\biggr)
        \nonumber \\ &
        -\biggl(
                27 \tilde{\rho} ^2+35 \tilde{\rho} ^3+\frac{\tilde{\rho} ^4}{12}+\frac{3 \tilde{\rho} ^5}{20}\biggr) \ln ^2(\tilde{\rho} )
        -\biggl(
                128 \tilde{\rho} ^{3/2}+\frac{9472}{15} \tilde{\rho} ^{5/2}+128 \tilde{\rho} ^{7/2}\biggr) \text{Li}_2\big(\sqrt{\tilde{\rho} }\big)
  \Biggr\}\,,
  \\
  Q_{2,0}^{(1)} &=
  C_F \Biggl\{
        \frac{302}{675}
        -\frac{21247 \tilde{\rho} }{900}
        -\frac{85018 \tilde{\rho} ^2}{225}
        +\frac{85018 \tilde{\rho} ^4}{225}
        +\frac{21247 \tilde{\rho} ^5}{900}
        -\frac{302 \tilde{\rho} ^6}{675}
        \nonumber \\ &
        +\ln (1-\tilde{\rho} ) \biggl[
                -\frac{112}{225}
                +\frac{763 \tilde{\rho} }{225}
                +\frac{280 \tilde{\rho} ^2}{9}
                -\frac{280 \tilde{\rho} ^4}{9}
                -\frac{763 \tilde{\rho} ^5}{225}
                +\frac{112 \tilde{\rho} ^6}{225}
                +\ln (\tilde{\rho} ) \biggl(
                        \frac{4}{15}
                        \nonumber \\ &
                        +\frac{4 \tilde{\rho} }{15}+\frac{476 \tilde{\rho} ^2}{3}+\frac{1400 \tilde{\rho} ^3}{3}+\frac{476 \tilde{\rho} ^4}{3}+\frac{4 \tilde{\rho} ^5}{15}+\frac{4 \tilde{\rho} ^6}{15}-32 \tilde{\rho} ^{3/2}-\frac{5408}{15} \tilde{\rho} ^{5/2}-\frac{5408}{15} \tilde{\rho} ^{7/2}
                        \nonumber \\ &
                        -32 \tilde{\rho} ^{9/2}\biggr)
        \biggr]
        +\ln (\tilde{\rho} ) \biggl[
                -\frac{68 \tilde{\rho} }{15}
                -\frac{933 \tilde{\rho} ^2}{5}
                -\frac{2132 \tilde{\rho} ^3}{5}
                -\frac{6997 \tilde{\rho} ^4}{45}
                -\frac{257 \tilde{\rho} ^5}{225}
                -\frac{112 \tilde{\rho} ^6}{225}
                \nonumber \\ &
                +\ln \big(1+\sqrt{\tilde{\rho} }\big)
\biggl(64 \tilde{\rho} ^{3/2}+\frac{10816}{15} \tilde{\rho} ^{5/2}+\frac{10816}{15} \tilde{\rho} ^{7/2}+64 \tilde{\rho} ^{9/2}\biggr)
        \biggr]
        +\text{Li}_2\big(\sqrt{\tilde{\rho} }\big)
\biggl(-128 \tilde{\rho} ^{3/2}
\nonumber \\ &
-\frac{21632}{15} \tilde{\rho} ^{5/2}-\frac{21632}{15} \tilde{\rho} ^{7/2}-128 \tilde{\rho} ^{9/2}\biggr)
        +\pi ^2 \biggl(
                -\frac{1}{15}-\frac{\tilde{\rho} }{15}-\frac{71 \tilde{\rho} ^2}{3}-\frac{640 \tilde{\rho} ^3}{9}-\frac{71 \tilde{\rho} ^4}{3}
                \nonumber \\ &
                -\frac{\tilde{\rho} ^5}{15}-\frac{\tilde{\rho} ^6}{15}+16 \tilde{\rho} ^{3/2}+\frac{2704}{15} \tilde{\rho} ^{5/2}+\frac{2704}{15} \tilde{\rho} ^{7/2}+16 \tilde{\rho} ^{9/2}\biggr)
        +\text{Li}_2(\tilde{\rho} ) \biggl(
                \frac{2}{5}+\frac{2 \tilde{\rho} }{5}+142 \tilde{\rho} ^2
                \nonumber \\ &
                +\frac{1280 \tilde{\rho} ^3}{3}+142 \tilde{\rho} ^4+\frac{2 \tilde{\rho} ^5}{5}+\frac{2 \tilde{\rho} ^6}{5}+32 \tilde{\rho} ^{3/2}+\frac{5408}{15} \tilde{\rho} ^{5/2}+\frac{5408}{15} \tilde{\rho} ^{7/2}+32 \tilde{\rho} ^{9/2}\biggr)
                \nonumber \\ &
                +\biggl(
                -36 \tilde{\rho} ^2-\frac{380 \tilde{\rho} ^3}{3}-\frac{155 \tilde{\rho} ^4}{3}-\frac{\tilde{\rho} ^5}{15}-\frac{\tilde{\rho} ^6}{15}\biggr) \ln ^2(\tilde{\rho} )
  \Biggr\}\,,
  \\
  Q_{3,0}^{(1)} &=
  C_F \Biggl\{
        \frac{1243}{5040}
        -\frac{724597 \tilde{\rho} }{35280}
        -\frac{38846267 \tilde{\rho} ^2}{58800}
        -\frac{174182263 \tilde{\rho} ^3}{176400}
        +\frac{174182263 \tilde{\rho} ^4}{176400}
        \nonumber \\ &
        +\frac{38846267 \tilde{\rho} ^5}{58800}
        +\frac{724597 \tilde{\rho} ^6}{35280}
        -\frac{1243 \tilde{\rho} ^7}{5040}
        +\ln ^2(\tilde{\rho} ) \biggl(
                -45 \tilde{\rho} ^2-300 \tilde{\rho} ^3-\frac{1359 \tilde{\rho} ^4}{4}-\frac{273 \tilde{\rho} ^5}{4}
                \nonumber \\ &
                -\frac{\tilde{\rho} ^6}{20}-\frac{\tilde{\rho} ^7}{28}\biggr)
        +\ln (1-\tilde{\rho} ) \biggl[
                -\frac{851}{2940}
                +\frac{763 \tilde{\rho} }{300}
                +\frac{917 \tilde{\rho} ^2}{20}
                +\frac{665 \tilde{\rho} ^3}{12}
                -\frac{665 \tilde{\rho} ^4}{12}
                -\frac{917 \tilde{\rho} ^5}{20}
                \nonumber \\ &
                -\frac{763 \tilde{\rho} ^6}{300}
                +\frac{851 \tilde{\rho} ^7}{2940}
                +\ln (\tilde{\rho} ) \biggl(
                        \frac{1}{7}+\frac{\tilde{\rho} }{5}+207 \tilde{\rho} ^2+1197 \tilde{\rho} ^3+1197 \tilde{\rho} ^4+207 \tilde{\rho} ^5+\frac{\tilde{\rho} ^6}{5}+\frac{\tilde{\rho} ^7}{7}
                        \nonumber \\ &
                        -32 \tilde{\rho} ^{3/2}-640 \tilde{\rho} ^{5/2}-\frac{51264}{35} \tilde{\rho} ^{7/2}-640 \tilde{\rho} ^{9/2}-32 \tilde{\rho} ^{11/2}\biggr)
        \biggr]
        +\ln (\tilde{\rho} ) \biggl[
                -\frac{101 \tilde{\rho} }{28}
                -\frac{38707 \tilde{\rho} ^2}{140}
                \nonumber \\ &
                -\frac{126527 \tilde{\rho} ^3}{105}
                -\frac{482833 \tilde{\rho} ^4}{420}
                -\frac{8072 \tilde{\rho} ^5}{35}
                -\frac{1117 \tilde{\rho} ^6}{1050}
                -\frac{851 \tilde{\rho} ^7}{2940}
                +\ln \big(1+\sqrt{\tilde{\rho} }\big)
\biggl(64 \tilde{\rho} ^{3/2}
\nonumber \\ &
+1280 \tilde{\rho} ^{5/2}+\frac{102528}{35} \tilde{\rho} ^{7/2}+1280 \tilde{\rho} ^{9/2}+64 \tilde{\rho} ^{11/2}\biggr)
        \biggr]
        +\text{Li}_2\big(\sqrt{\tilde{\rho} }\big)
\biggl(-128 \tilde{\rho} ^{3/2}-2560 \tilde{\rho} ^{5/2}
\nonumber \\ &
-\frac{205056}{35} \tilde{\rho} ^{7/2}-2560 \tilde{\rho} ^{9/2}-128 \tilde{\rho} ^{11/2}\biggr)
        +\pi ^2 \biggl(
                -\frac{1}{28}-\frac{\tilde{\rho} }{20}-\frac{125 \tilde{\rho} ^2}{4}-\frac{743 \tilde{\rho} ^3}{4}
                \nonumber \\ &
                -\frac{743 \tilde{\rho} ^4}{4}-\frac{125 \tilde{\rho} ^5}{4}-\frac{\tilde{\rho} ^6}{20}-\frac{\tilde{\rho} ^7}{28}+16 \tilde{\rho} ^{3/2}+320 \tilde{\rho} ^{5/2}+\frac{25632}{35} \tilde{\rho} ^{7/2}+320 \tilde{\rho} ^{9/2}+16 \tilde{\rho} ^{11/2}\biggr)
                \nonumber \\ &
                +\text{Li}_2(\tilde{\rho} ) \biggl(
                \frac{3}{14}+\frac{3 \tilde{\rho} }{10}+\frac{375 \tilde{\rho} ^2}{2}+\frac{2229 \tilde{\rho} ^3}{2}+\frac{2229 \tilde{\rho} ^4}{2}+\frac{375 \tilde{\rho} ^5}{2}+\frac{3 \tilde{\rho} ^6}{10}+\frac{3 \tilde{\rho} ^7}{14}+32 \tilde{\rho} ^{3/2}
                \nonumber \\ &
                +640 \tilde{\rho} ^{5/2}+\frac{51264}{35} \tilde{\rho} ^{7/2}+640 \tilde{\rho} ^{9/2}+32 \tilde{\rho} ^{11/2}\biggr)
\Biggr\}\,,
  \\
  Q_{4,0}^{(1)} &=
  C_F \Biggl\{
        \frac{1429}{9408}
        -\frac{970901 \tilde{\rho} }{52920}
        -\frac{175592549 \tilde{\rho} ^2}{176400}
        -\frac{313385041 \tilde{\rho} ^3}{88200}
        +\frac{313385041 \tilde{\rho} ^5}{88200}
        \nonumber \\ &
        +\frac{175592549 \tilde{\rho} ^6}{176400}
        +\frac{970901 \tilde{\rho} ^7}{52920}
        -\frac{1429 \tilde{\rho} ^8}{9408}
        +\ln ^2(\tilde{\rho} ) \biggl(
                -54 \tilde{\rho} ^2-580 \tilde{\rho} ^3-\frac{3773 \tilde{\rho} ^4}{3}
                \nonumber \\ &
                -\frac{3536 \tilde{\rho} ^5}{5}-\frac{424 \tilde{\rho} ^6}{5}-\frac{4 \tilde{\rho} ^7}{105}-\frac{3 \tilde{\rho} ^8}{140}\biggr)
        +\ln (1-\tilde{\rho} ) \biggl[
                -\frac{5449}{29400}
                +\frac{7508 \tilde{\rho} }{3675}
                +\frac{4592 \tilde{\rho} ^2}{75}
                \nonumber \\ &
                +\frac{12628 \tilde{\rho} ^3}{75}
                -\frac{12628 \tilde{\rho} ^5}{75}
                -\frac{4592 \tilde{\rho} ^6}{75}
                -\frac{7508 \tilde{\rho} ^7}{3675}
                +\frac{5449 \tilde{\rho} ^8}{29400}
                +\ln (\tilde{\rho} ) \biggl(
                        \frac{3}{35}+\frac{16 \tilde{\rho} }{105}+\frac{1276 \tilde{\rho} ^2}{5}
                        \nonumber \\ &
                        +\frac{12144 \tilde{\rho} ^3}{5}+4774 \tilde{\rho} ^4+\frac{12144 \tilde{\rho} ^5}{5}+\frac{1276 \tilde{\rho} ^6}{5}+\frac{16 \tilde{\rho} ^7}{105}+\frac{3 \tilde{\rho} ^8}{35}-32 \tilde{\rho} ^{3/2}-\frac{14944}{15} \tilde{\rho} ^{5/2}
                        \nonumber \\ &
                        -\frac{141504}{35} \tilde{\rho} ^{7/2}-\frac{141504}{35} \tilde{\rho} ^{9/2}-\frac{14944}{15} \tilde{\rho} ^{11/2}-32 \tilde{\rho} ^{13/2}\biggr)
        \biggr]
        +\ln (\tilde{\rho} ) \biggl[
                -3 \tilde{\rho} 
                -\frac{39449 \tilde{\rho} ^2}{105}
                \nonumber \\ &
                -\frac{280199 \tilde{\rho} ^3}{105}
                -\frac{2987011 \tilde{\rho} ^4}{630}
                -\frac{437533 \tilde{\rho} ^5}{175}
                -\frac{165101 \tilde{\rho} ^6}{525}
                -\frac{3517 \tilde{\rho} ^7}{3675}
                -\frac{5449 \tilde{\rho} ^8}{29400}
                \nonumber \\ &
                +\ln \big(1+\sqrt{\tilde{\rho} }\big)
\biggl(64 \tilde{\rho} ^{3/2}+\frac{29888}{15} \tilde{\rho} ^{5/2}+\frac{283008}{35} \tilde{\rho} ^{7/2}+\frac{283008}{35} \tilde{\rho} ^{9/2}+\frac{29888}{15} \tilde{\rho} ^{11/2}
\nonumber \\ &
+64 \tilde{\rho} ^{13/2}\biggr)
        \biggr]
        +\text{Li}_2\big(\sqrt{\tilde{\rho} }\big)
\biggl(-128 \tilde{\rho} ^{3/2}-\frac{59776}{15} \tilde{\rho} ^{5/2}-\frac{566016}{35} \tilde{\rho} ^{7/2}-\frac{566016}{35} \tilde{\rho} ^{9/2}
\nonumber \\ &
-\frac{59776}{15} \tilde{\rho} ^{11/2}-128 \tilde{\rho} ^{13/2}\biggr)
        +\pi ^2 \biggl(
                -\frac{3}{140}-\frac{4 \tilde{\rho} }{105}-\frac{194 \tilde{\rho} ^2}{5}-\frac{5708 \tilde{\rho} ^3}{15}-\frac{6776 \tilde{\rho} ^4}{9}
                \nonumber \\ &
                -\frac{5708 \tilde{\rho} ^5}{15}-\frac{194 \tilde{\rho} ^6}{5}-\frac{4 \tilde{\rho} ^7}{105}-\frac{3 \tilde{\rho} ^8}{140}+16 \tilde{\rho} ^{3/2}+\frac{7472}{15} \tilde{\rho} ^{5/2}+\frac{70752}{35} \tilde{\rho} ^{7/2}+\frac{70752}{35} \tilde{\rho} ^{9/2}
                \nonumber \\ &
                +\frac{7472}{15} \tilde{\rho} ^{11/2}+16 \tilde{\rho} ^{13/2}\biggr)
        +\text{Li}_2(\tilde{\rho} ) \biggl(
                \frac{9}{70}+\frac{8 \tilde{\rho} }{35}+\frac{1164 \tilde{\rho} ^2}{5}+\frac{11416 \tilde{\rho} ^3}{5}+\frac{13552 \tilde{\rho} ^4}{3}
                \nonumber \\ &
                +\frac{11416 \tilde{\rho} ^5}{5}+\frac{1164 \tilde{\rho} ^6}{5}+\frac{8 \tilde{\rho} ^7}{35}+\frac{9 \tilde{\rho} ^8}{70}+32 \tilde{\rho} ^{3/2}+\frac{14944}{15} \tilde{\rho} ^{5/2}+\frac{141504}{35} \tilde{\rho} ^{7/2}+\frac{141504}{35} \tilde{\rho} ^{9/2}
                \nonumber \\ &
                +\frac{14944}{15} \tilde{\rho} ^{11/2}+32 \tilde{\rho} ^{13/2}\biggr)
\Biggr\}\,.
\end{align}

The analytic results for power-suppressed contributions
$Q_{i,0, \mu_G}^{(0)}$, $Q_{i,0, \mu_G}^{(1)}$ and $Q_{i,0, \rho_{LS}}^{(1)}$
are given by
\begin{align}
  Q_{0,0, \mu_G}^{(0)} &= -\frac{5 \tilde{\rho} ^4}{2}+12 \tilde{\rho} ^3-12 \tilde{\rho} ^2-6 \tilde{\rho} ^2 \ln (\tilde{\rho} )+4 \tilde{\rho}-\frac{3}{2}\,,
  \\
  Q_{1,0, \mu_G}^{(0)} &= \tilde{\rho}  \left(-9 \tilde{\rho} ^2+3 \tilde{\rho} -4\right) \ln (\tilde{\rho} )+\frac{1}{12} \left(-9 \tilde{\rho}^5+79 \tilde{\rho} ^4-48 \tilde{\rho} ^3-7 \tilde{\rho} -15\right)\,,
  \\
  Q_{2,0, \mu_G}^{(0)} &= -\frac{1}{3} (\tilde{\rho} -1)^2 \left(\tilde{\rho} ^4-12 \tilde{\rho} ^3-36 \tilde{\rho} ^2+44 \tilde{\rho} +12 (3 \tilde{\rho}+2) \tilde{\rho}  \ln (\tilde{\rho} )+3\right)\,,
  \\
  Q_{3,0, \mu_G}^{(0)} &= \frac{1}{140} \left(-25 \tilde{\rho} ^7+511 \tilde{\rho} ^6+1715 \tilde{\rho} ^5-18725 \tilde{\rho} ^4+11025 \tilde{\rho}^3+9625 \tilde{\rho} ^2-4011 \tilde{\rho} -115\right)
  \nonumber \\ &
    -3 \tilde{\rho}  \left(5 \tilde{\rho} ^4-15 \tilde{\rho} ^3-35\tilde{\rho} ^2+5 \tilde{\rho} +4\right) \ln (\tilde{\rho} )\,,
  \\
  Q_{4,0, \mu_G}^{(0)} &= \frac{1}{420} \left(-45 \tilde{\rho} ^8+1264 \tilde{\rho} ^7+9156 \tilde{\rho} ^6-140784 \tilde{\rho} ^5-81200\tilde{\rho} ^4+216720 \tilde{\rho} ^3+15036 \tilde{\rho} ^2 \right.
  \nonumber \\ &
    \left. -19856 \tilde{\rho} -291\right)-2 \tilde{\rho}  \left(9\tilde{\rho} ^5-48 \tilde{\rho} ^4-245 \tilde{\rho} ^3-120 \tilde{\rho} ^2+33 \tilde{\rho} +8\right) \ln (\tilde{\rho} )\,,
\end{align}
\begin{align}
  Q_{0,0, \mu_G}^{(1)} &= C_A 
  \Biggl\{
    -\frac{1}{36}
    -\frac{3805 \tilde{\rho} }{216}
    +\frac{2431 \tilde{\rho} ^2}{216}
    +\frac{415 \tilde{\rho} ^3}{72}
    +\frac{5 \tilde{\rho} ^4}{8}
    +\ln \big(1-\sqrt{\tilde{\rho} }\big)
    \biggl[
      -\frac{5}{108}
      -\frac{5}{36 \tilde{\rho} }
      \nonumber \\ &
      +2 \tilde{\rho} 
      +\frac{8 \tilde{\rho} ^2}{9}
      -\frac{265 \tilde{\rho} ^3}{108}
      -\frac{\tilde{\rho} ^4}{4}
      +\ln (\tilde{\rho} ) 
      \biggl(
        -\frac{17}{9}+5 \tilde{\rho} +\frac{5 \tilde{\rho} ^2}{3}+\frac{37 \tilde{\rho} ^3}{9}+\frac{16}{3} \sqrt{\tilde{\rho} }-\frac{176}{9} \tilde{\rho} ^{3/2}
      \biggr)
    \biggr]
    \nonumber \\ &
    +\ln (\tilde{\rho} ) \biggl[
            -\frac{131 \tilde{\rho} }{18}
            -12 \tilde{\rho} ^2
            -\frac{131 \tilde{\rho} ^3}{108}
            +\frac{\tilde{\rho} ^4}{4}
            +\ln (1-\tilde{\rho} ) 
            \big(
              \frac{16}{9}-6 \tilde{\rho} +\frac{26 \tilde{\rho} ^2}{3}-\frac{56 \tilde{\rho} ^3}{9}
              \nonumber \\ &
              -\frac{8}{3} \sqrt{\tilde{\rho} }+\frac{88}{9} \tilde{\rho} ^{3/2}
            \big)
    \biggr]
    +\ln\left(\frac{\mu_s}{m_b}\right)
    \biggl(-\frac{3}{4}
            +2 \tilde{\rho} 
            -6 \tilde{\rho} ^2
            +6 \tilde{\rho} ^3
            -\frac{5 \tilde{\rho} ^4}{4}
            -3 \tilde{\rho} ^2 \ln (\tilde{\rho} )
    \biggr)
    \nonumber \\ &
    +\ln \big(1+\sqrt{\tilde{\rho} }\big)
    \biggl[ -\frac{5}{108}
            -\frac{5}{36 \tilde{\rho} }
            +2 \tilde{\rho} 
            +\frac{8 \tilde{\rho} ^2}{9}
            -\frac{265 \tilde{\rho} ^3}{108}
            -\frac{\tilde{\rho} ^4}{4}
            +\biggl(-\frac{17}{9}+5 \tilde{\rho} +\frac{5 \tilde{\rho} ^2}{3}
            \nonumber \\ &
            +\frac{37 \tilde{\rho} ^3}{9}\biggr) \ln (\tilde{\rho} )
    \biggr]
    +\pi ^2 \biggl(
            -\frac{1}{18}+
            \frac{\tilde{\rho} }{3}
            -\frac{4 \tilde{\rho} ^2}{3}
            +\frac{11 \tilde{\rho} ^3}{18}
            -\frac{4}{3} \sqrt{\tilde{\rho} }
            +\frac{44}{9} \tilde{\rho} ^{3/2}
    \biggr)
    \nonumber \\ &
    +\text{Li}_2(\tilde{\rho} ) \biggl(
            \frac{1}{3}-2 \tilde{\rho} +8 \tilde{\rho} ^2-\frac{11 \tilde{\rho} ^3}{3}-\frac{8}{3} \sqrt{\tilde{\rho} }+\frac{88}{9} \tilde{\rho} ^{3/2}\biggr)
    +\biggl(
            \frac{\tilde{\rho} }{2}-\frac{11 \tilde{\rho} ^2}{6}+\frac{5 \tilde{\rho} ^3}{18}\biggr) \ln ^2(\tilde{\rho} )
    \nonumber \\ &
    +\biggl(
            \frac{32}{3} \sqrt{\tilde{\rho} }-\frac{352}{9} \tilde{\rho} ^{3/2}\biggr) \text{Li}_2\big(
            \sqrt{\tilde{\rho} }\big)
\Biggr\}
+C_F \Biggl\{
    -\frac{215}{144}
    +\frac{3733 \tilde{\rho} }{108}
    -\frac{1360 \tilde{\rho} ^2}{27}
    +\frac{985 \tilde{\rho} ^3}{36}
    \nonumber \\ &
    -\frac{161 \tilde{\rho} ^4}{16}
    +\ln (\tilde{\rho} ) \biggl[
            \frac{125 \tilde{\rho} }{9}
            -\frac{109 \tilde{\rho} ^2}{18}
            +\frac{667 \tilde{\rho} ^3}{27}
            -\frac{65 \tilde{\rho} ^4}{12}
            +\ln (1-\tilde{\rho} ) \biggl(
                    -\frac{14}{9}+24 \tilde{\rho} 
                    \nonumber \\ &
                    -\frac{74 \tilde{\rho} ^2}{3}-\frac{152 \tilde{\rho} ^3}{9}+10 \tilde{\rho} ^4+\frac{32}{3} \sqrt{\tilde{\rho} }-\frac{128}{9} \tilde{\rho} ^{3/2}\biggr)
    \biggr]
    +\ln \big(1-\sqrt{\tilde{\rho} }\big)
\biggl[\frac{31}{108}
            -\frac{1}{9 \tilde{\rho} }
            -\frac{26 \tilde{\rho} }{3}
            \nonumber \\ &
            +\frac{226 \tilde{\rho} ^2}{9}
            -\frac{595 \tilde{\rho} ^3}{27}
            +\frac{65 \tilde{\rho} ^4}{12}
            +\ln (\tilde{\rho} ) \biggl(
                    \frac{25}{9}-18 \tilde{\rho} +\frac{58 \tilde{\rho} ^2}{3}+\frac{58 \tilde{\rho} ^3}{9}-5 \tilde{\rho} ^4-\frac{64}{3} \sqrt{\tilde{\rho} }
                    \nonumber \\ &
                    +\frac{256}{9} \tilde{\rho} ^{3/2}\biggr)
    \biggr]
    +\ln \big(1+\sqrt{\tilde{\rho} }\big)
\biggl[\frac{31}{108}
            -\frac{1}{9 \tilde{\rho} }
            -\frac{26 \tilde{\rho} }{3}
            +\frac{226 \tilde{\rho} ^2}{9}
            -\frac{595 \tilde{\rho} ^3}{27}
            +\frac{65 \tilde{\rho} ^4}{12}
            \nonumber \\ &
            +\biggl(
                    \frac{25}{9}-18 \tilde{\rho} +\frac{58 \tilde{\rho} ^2}{3}+\frac{58 \tilde{\rho} ^3}{9}-5 \tilde{\rho} ^4\biggr) \ln (\tilde{\rho} )
    \biggr]
    +\text{Li}_2(\tilde{\rho} ) \biggl(
            \frac{5}{6}+12 \tilde{\rho} -12 \tilde{\rho} ^2
            \nonumber \\ &
            -\frac{44 \tilde{\rho} ^3}{3}+\frac{15 \tilde{\rho} ^4}{2}+\frac{32}{3} \sqrt{\tilde{\rho} }-\frac{128}{9} \tilde{\rho} ^{3/2}\biggr)
    +\pi ^2 \biggl(
            -\frac{5}{36}-2 \tilde{\rho} +2 \tilde{\rho} ^2+\frac{22 \tilde{\rho} ^3}{9}-\frac{5 \tilde{\rho} ^4}{4}
            \nonumber \\ &
            +\frac{16}{3} \sqrt{\tilde{\rho} }-\frac{64}{9} \tilde{\rho} ^{3/2}\biggr)
    +\biggl(
            -2 \tilde{\rho} -\frac{5 \tilde{\rho} ^2}{3}+\frac{28 \tilde{\rho} ^3}{9}-\frac{5 \tilde{\rho} ^4}{4}\biggr) \ln ^2(\tilde{\rho} )
            \nonumber \\ &
    +\biggl(
            -\frac{128}{3} \sqrt{\tilde{\rho} }+\frac{512}{9} \tilde{\rho} ^{3/2}\biggr) \text{Li}_2\big(
            \sqrt{\tilde{\rho} }\big)
\Biggr\}\,,
  \\
  Q_{1,0, \mu_G}^{(1)} &= C_A \Biggl\{
    \frac{139}{180}
    -\frac{29779 \tilde{\rho} }{2160}
    -\frac{8398 \tilde{\rho} ^2}{135}
    +\frac{74623 \tilde{\rho} ^3}{1080}
    +\frac{1069 \tilde{\rho} ^4}{180}
    +\frac{3 \tilde{\rho} ^5}{16}
    \nonumber \\ &
    +\ln \big(1-\sqrt{\tilde{\rho} }\big)
\biggl[-\frac{67}{540}
            -\frac{17}{360 \tilde{\rho} }
            +\frac{1453 \tilde{\rho} }{216}
            +\frac{11 \tilde{\rho} ^2}{9}
            -\frac{1145 \tilde{\rho} ^3}{216}
            -\frac{1297 \tilde{\rho} ^4}{540}
            -\frac{3 \tilde{\rho} ^5}{40}
            \nonumber \\ &
            +\ln (\tilde{\rho} ) \biggl(
                    -\frac{5}{9}+\frac{47 \tilde{\rho} }{9}+\frac{50 \tilde{\rho} ^2}{3}+\frac{91 \tilde{\rho} ^3}{9}+\frac{17 \tilde{\rho} ^4}{9}+\frac{16}{3} \sqrt{\tilde{\rho} }+\frac{64}{9} \tilde{\rho} ^{3/2}-\frac{880}{9} \tilde{\rho} ^{5/2}\biggr)
    \biggr]
    \nonumber \\ &
    +\ln (\tilde{\rho} ) \biggl[
            \frac{29 \tilde{\rho} }{36}
            -\frac{383 \tilde{\rho} ^2}{9}
            -\frac{7723 \tilde{\rho} ^3}{216}
            +\frac{43 \tilde{\rho} ^4}{135}
            +\frac{3 \tilde{\rho} ^5}{40}
            +\ln (1-\tilde{\rho} ) \biggl(
                    \frac{10}{9}-\frac{157 \tilde{\rho} }{9}+\frac{50 \tilde{\rho} ^2}{3}
                    \nonumber \\ &
                    +\frac{115 \tilde{\rho} ^3}{9}-\frac{34 \tilde{\rho} ^4}{9}-\frac{8}{3} \sqrt{\tilde{\rho} }-\frac{32}{9} \tilde{\rho} ^{3/2}+\frac{440}{9} \tilde{\rho} ^{5/2}\biggr)
    \biggr]
    +\ln \left(\frac{\mu_s}{m_b}\right)
\biggl[-\frac{5}{8}
            -\frac{7 \tilde{\rho} }{24}
            -2 \tilde{\rho} ^3
            \nonumber \\ &
            +\frac{79 \tilde{\rho} ^4}{24}
            -\frac{3 \tilde{\rho} ^5}{8}
            +\biggl(
                    -2 \tilde{\rho} +\frac{3 \tilde{\rho} ^2}{2}-
                    \frac{9 \tilde{\rho} ^3}{2}\biggr) \ln (\tilde{\rho} )
    \biggr]
    +\ln \big(1+\sqrt{\tilde{\rho} }\big)
\biggl[-\frac{67}{540}
            -\frac{17}{360 \tilde{\rho} }
            \nonumber \\ &
            +\frac{1453 \tilde{\rho} }{216}
            +\frac{11 \tilde{\rho} ^2}{9}
            -\frac{1145 \tilde{\rho} ^3}{216}
            -\frac{1297 \tilde{\rho} ^4}{540}
            -\frac{3 \tilde{\rho} ^5}{40}
            +\biggl(
                    -\frac{5}{9}+\frac{47 \tilde{\rho} }{9}+\frac{50 \tilde{\rho} ^2}{3}+\frac{91 \tilde{\rho} ^3}{9}
                    \nonumber \\ &
                    +\frac{17 \tilde{\rho} ^4}{9}\biggr) \ln (\tilde{\rho} )
    \biggr]
    +\pi ^2 \biggl(
            -\frac{5}{36}+\frac{89 \tilde{\rho} }{36}-\frac{25 \tilde{\rho} ^2}{6}-\frac{107 \tilde{\rho} ^3}{36}+\frac{17 \tilde{\rho} ^4}{36}-\frac{4}{3} \sqrt{\tilde{\rho} }
            -\frac{16}{9} \tilde{\rho} ^{3/2}
            \nonumber \\ &
            +\frac{220}{9} \tilde{\rho} ^{5/2}\biggr)
    +\text{Li}_2(\tilde{\rho} ) \biggl(
            \frac{5}{6}-\frac{89 \tilde{\rho} }{6}+25 \tilde{\rho} ^2+\frac{107 \tilde{\rho} ^3}{6}-\frac{17 \tilde{\rho} ^4}{6}
            -\frac{8}{3} \sqrt{\tilde{\rho} }
            -\frac{32}{9} \tilde{\rho} ^{3/2}
            \nonumber \\ &
            +\frac{440}{9} \tilde{\rho} ^{5/2}\biggr)
    +\biggl(
            \frac{11 \tilde{\rho} }{4}-\frac{16 \tilde{\rho} ^2}{3}-\frac{103 \tilde{\rho} ^3}{18}+\frac{17 \tilde{\rho} ^4}{36}\biggr) \ln ^2(\tilde{\rho} )
    +\biggl(
            \frac{32}{3} \sqrt{\tilde{\rho} }+\frac{128}{9} \tilde{\rho} ^{3/2}
            \nonumber \\ &
            -\frac{1760}{9} \tilde{\rho} ^{5/2}\biggr) \text{Li}_2\big(
            \sqrt{\tilde{\rho} }\big)
\Biggr\}
+C_F \Biggl\{
    -\frac{2771}{720}
    +\frac{75529 \tilde{\rho} }{2160}
    +\frac{70027 \tilde{\rho} ^2}{1080}
    -\frac{7597 \tilde{\rho} ^3}{72}
    +\frac{9121 \tilde{\rho} ^4}{720}
    \nonumber \\ &
    -\frac{249 \tilde{\rho} ^5}{80}
    +\ln (\tilde{\rho} ) \biggl[
            -\frac{91 \tilde{\rho} }{72}
            +\frac{317 \tilde{\rho} ^2}{4}
            +\frac{3815 \tilde{\rho} ^3}{108}
            +\frac{2527 \tilde{\rho} ^4}{180}
            -\frac{241 \tilde{\rho} ^5}{120}
            +\ln (1-\tilde{\rho} ) \biggl(
                    -\frac{13}{9}
                    \nonumber \\ &
                    +\frac{109 \tilde{\rho} }{3}-\frac{197 \tilde{\rho} ^2}{3}-\frac{463 \tilde{\rho} ^3}{9}-9 \tilde{\rho} ^4+3 \tilde{\rho} ^5+\frac{32}{3} \sqrt{\tilde{\rho} }+\frac{32}{9} \tilde{\rho} ^{3/2}-\frac{256}{3} \tilde{\rho} ^{5/2}\biggr)
    \biggr]
    \nonumber \\ &
    +\ln \big(1-\sqrt{\tilde{\rho} }\big)
\biggl[\frac{1511}{1080}
            -\frac{2}{45 \tilde{\rho} }
            -\frac{623 \tilde{\rho} }{72}
            +\frac{109 \tilde{\rho} ^2}{9}
            +\frac{187 \tilde{\rho} ^3}{27}
            -\frac{4949 \tilde{\rho} ^4}{360}
            +\frac{241 \tilde{\rho} ^5}{120}
            \nonumber \\ &
            +\ln (\tilde{\rho} ) \biggl(
                    \frac{13}{18}-\frac{13 \tilde{\rho} }{6}-\frac{38 \tilde{\rho} ^2}{3}+\frac{254 \tilde{\rho} ^3}{9}+\frac{9 \tilde{\rho} ^4}{2}-\frac{3 \tilde{\rho} ^5}{2}-\frac{64}{3} \sqrt{\tilde{\rho} }-\frac{64}{9} \tilde{\rho} ^{3/2}+\frac{512}{3} \tilde{\rho} ^{5/2}\biggr)
    \biggr]
    \nonumber \\ &
    +\ln \big(1+\sqrt{\tilde{\rho} }\big)
\biggl[\frac{1511}{1080}
            -\frac{2}{45 \tilde{\rho} }
            -\frac{623 \tilde{\rho} }{72}
            +\frac{109 \tilde{\rho} ^2}{9}
            +\frac{187 \tilde{\rho} ^3}{27}
            -\frac{4949 \tilde{\rho} ^4}{360}
            +\frac{241 \tilde{\rho} ^5}{120}
            +\biggl(
                    \frac{13}{18}
                    \nonumber \\ &
                    -\frac{13 \tilde{\rho} }{6}-\frac{38 \tilde{\rho} ^2}{3}+\frac{254 \tilde{\rho} ^3}{9}+\frac{9 \tilde{\rho} ^4}{2}-\frac{3 \tilde{\rho} ^5}{2}\biggr) \ln (\tilde{\rho} )
    \biggr]
    +\text{Li}_2(\tilde{\rho} ) \biggl(
            -\frac{13}{12}+\frac{141 \tilde{\rho} }{4}-72 \tilde{\rho} ^2
            \nonumber \\ &
            -\frac{112 \tilde{\rho} ^3}{3}-\frac{27 \tilde{\rho} ^4}{4}+\frac{9 \tilde{\rho} ^5}{4}+\frac{32}{3} \sqrt{\tilde{\rho} }+\frac{32}{9} \tilde{\rho} ^{3/2}-\frac{256}{3} \tilde{\rho} ^{5/2}\biggr)
    +\pi ^2 \biggl(
            \frac{13}{72}-\frac{47 \tilde{\rho} }{8}+12 \tilde{\rho} ^2
            \nonumber \\ &
            +\frac{56 \tilde{\rho} ^3}{9}+\frac{9 \tilde{\rho} ^4}{8}-\frac{3 \tilde{\rho} ^5}{8}+\frac{16}{3} \sqrt{\tilde{\rho} }+\frac{16}{9} \tilde{\rho} ^{3/2}-\frac{128}{3} \tilde{\rho} ^{5/2}\biggr)
    +\biggl(
            -7 \tilde{\rho} +\frac{83 \tilde{\rho} ^2}{6}+\frac{37 \tilde{\rho} ^3}{18}
            \nonumber \\ &
            +\frac{9 \tilde{\rho} ^4}{8}-\frac{3 \tilde{\rho} ^5}{8}\biggr) \ln ^2(\tilde{\rho} )
    +\biggl(
            -\frac{128}{3} \sqrt{\tilde{\rho} }-\frac{128}{9} \tilde{\rho} ^{3/2}+\frac{1024}{3} \tilde{\rho} ^{5/2}\biggr) \text{Li}_2\big(
            \sqrt{\tilde{\rho} }\big)
  \Biggr\}\,,
  \\
  Q_{2,0, \mu_G}^{(1)} &= C_F \Biggl\{
    -\frac{293}{90}
    -\frac{237923 \tilde{\rho} }{5400}
    +\frac{362632 \tilde{\rho} ^2}{675}
    -\frac{417839 \tilde{\rho} ^3}{1350}
    -\frac{127349 \tilde{\rho} ^4}{675}
    +\frac{10451 \tilde{\rho} ^5}{1080}
    \nonumber \\ &
    -\frac{383 \tilde{\rho} ^6}{270}
    +\ln ^2(\tilde{\rho} ) \biggl(
            -12 \tilde{\rho} +\frac{29 \tilde{\rho} ^2}{3}+\frac{833 \tilde{\rho} ^3}{9}+\frac{20 \tilde{\rho} ^4}{9}+\frac{37 \tilde{\rho} ^5}{45}-\frac{\tilde{\rho} ^6}{6}\biggr)
            \nonumber \\ &
    +\ln \big(1+\sqrt{\tilde{\rho} }\big)
\biggl[\frac{2053}{1350}
            -\frac{1}{45 \tilde{\rho} }
            -\frac{338 \tilde{\rho} }{135}
            -\frac{550 \tilde{\rho} ^2}{27}
            +\frac{1135 \tilde{\rho} ^3}{27}
            -\frac{2969 \tilde{\rho} ^4}{270}
            -\frac{7214 \tilde{\rho} ^5}{675}
            \nonumber \\ &
            +\frac{46 \tilde{\rho} ^6}{45}
            +\ln (\tilde{\rho} ) \biggl(
                    \frac{28}{45}+\frac{44 \tilde{\rho} }{9}-\frac{250 \tilde{\rho} ^2}{9}-\frac{40 \tilde{\rho} ^3}{9}+\frac{368 \tilde{\rho} ^4}{9}+\frac{148 \tilde{\rho} ^5}{45}-\frac{2 \tilde{\rho} ^6}{3}\biggr)
    \biggr]
    \nonumber \\ &
    +\ln (\tilde{\rho} ) \biggl[
            -\frac{1573 \tilde{\rho} }{45}
            +\frac{17197 \tilde{\rho} ^2}{90}
            +\frac{51568 \tilde{\rho} ^3}{135}
            +\frac{19901 \tilde{\rho} ^4}{270}
            +\frac{14203 \tilde{\rho} ^5}{1350}
            -\frac{46 \tilde{\rho} ^6}{45}
            \nonumber \\ &
            +\ln (1-\tilde{\rho} ) \biggl(
                    -\frac{56}{45}+\frac{488 \tilde{\rho} }{9}-\frac{520 \tilde{\rho} ^2}{9}-\frac{3544 \tilde{\rho} ^3}{9}-\frac{664 \tilde{\rho} ^4}{9}-\frac{296 \tilde{\rho} ^5}{45}+
                    \frac{4 \tilde{\rho} ^6}{3}+\frac{32}{3} \sqrt{\tilde{\rho} }
                    \nonumber \\ &
                    +\frac{704}{9} \tilde{\rho} ^{3/2}-\frac{15776}{45} \tilde{\rho} ^{5/2}-\frac{1792}{9} \tilde{\rho} ^{7/2}\biggr)
    \biggr]
    +\ln \big(1-\sqrt{\tilde{\rho} }\big)
\biggl[\frac{2053}{1350}
            -\frac{1}{45 \tilde{\rho} }
            -\frac{338 \tilde{\rho} }{135}
            \nonumber \\ &
            -\frac{550 \tilde{\rho} ^2}{27}
            +\frac{1135 \tilde{\rho} ^3}{27}
            -\frac{2969 \tilde{\rho} ^4}{270}
            -\frac{7214 \tilde{\rho} ^5}{675}
            +\frac{46 \tilde{\rho} ^6}{45}
            +\ln (\tilde{\rho} ) \biggl(
                    \frac{28}{45}+\frac{44 \tilde{\rho} }{9}-\frac{250 \tilde{\rho} ^2}{9}
                    \nonumber \\ &
                    -\frac{40 \tilde{\rho} ^3}{9}+\frac{368 \tilde{\rho} ^4}{9}+\frac{148 \tilde{\rho} ^5}{45}-\frac{2 \tilde{\rho} ^6}{3}-\frac{64}{3} \sqrt{\tilde{\rho} }-\frac{1408}{9} \tilde{\rho} ^{3/2}+\frac{31552}{45} \tilde{\rho} ^{5/2}+\frac{3584}{9} \tilde{\rho} ^{7/2}\biggr)
    \biggr]
    \nonumber \\ &
    +\text{Li}_2(\tilde{\rho} ) \biggl(
            -\frac{14}{15}+\frac{170 \tilde{\rho} }{3}-\frac{215 \tilde{\rho} ^2}{3}-396 \tilde{\rho} ^3-\frac{160 \tilde{\rho} ^4}{3}-\frac{74 \tilde{\rho} ^5}{15}+\tilde{\rho} ^6+\frac{32}{3} \sqrt{\tilde{\rho} }+\frac{704}{9} \tilde{\rho} ^{3/2}
            \nonumber \\ &
            -\frac{15776}{45} \tilde{\rho} ^{5/2}-\frac{1792}{9} \tilde{\rho} ^{7/2}\biggr)
    +\pi ^2 \biggl(
            \frac{7}{45}-\frac{85 \tilde{\rho} }{9}+\frac{215 \tilde{\rho} ^2}{18}+66 \tilde{\rho} ^3+\frac{80 \tilde{\rho} ^4}{9}+\frac{37 \tilde{\rho} ^5}{45}-
            \frac{\tilde{\rho} ^6}{6}
            \nonumber \\ &
            +\frac{16}{3} \sqrt{\tilde{\rho} }+\frac{352}{9} \tilde{\rho} ^{3/2}-\frac{7888}{45} \tilde{\rho} ^{5/2}-\frac{896}{9} \tilde{\rho} ^{7/2}\biggr)
    +\text{Li}_2\big(\sqrt{\tilde{\rho} }\big)
\biggl(-\frac{128}{3} \sqrt{\tilde{\rho} }-\frac{2816}{9} \tilde{\rho} ^{3/2}
\nonumber \\ &
+\frac{63104}{45} \tilde{\rho} ^{5/2}+\frac{7168}{9} \tilde{\rho} ^{7/2}\biggr)
\Biggr\}
+C_A \Biggl\{
    \frac{683}{1800}
    +\frac{22571 \tilde{\rho} }{2700}
    -\frac{322361 \tilde{\rho} ^2}{1350}
    +\frac{161147 \tilde{\rho} ^3}{2700}
    \nonumber \\ &
    +\frac{297611 \tilde{\rho} ^4}{1800}
    +\frac{247 \tilde{\rho} ^5}{50}
    +\frac{\tilde{\rho} ^6}{12}
    +\ln \big(1-\sqrt{\tilde{\rho} }\big)
\biggl[-\frac{319}{2700}
            -\frac{1}{45 \tilde{\rho} }
            +\frac{6143 \tilde{\rho} }{540}
            +\frac{295 \tilde{\rho} ^2}{18}
            \nonumber \\ &
            -\frac{365 \tilde{\rho} ^3}{27}
            -\frac{6413 \tilde{\rho} ^4}{540}
            -\frac{659 \tilde{\rho} ^5}{300}
            -\frac{\tilde{\rho} ^6}{30}
            +\ln (\tilde{\rho} ) \biggl(
                    -\frac{17}{45}+\frac{67 \tilde{\rho} }{9}+\frac{128 \tilde{\rho} ^2}{3}+\frac{404 \tilde{\rho} ^3}{9}+\frac{127 \tilde{\rho} ^4}{9}
                    \nonumber \\ &
                    +\frac{7 \tilde{\rho} ^5}{5}+\frac{16}{3} \sqrt{\tilde{\rho} }+\frac{560}{9} \tilde{\rho} ^{3/2}-\frac{12208}{45} \tilde{\rho} ^{5/2}-240 \tilde{\rho} ^{7/2}\biggr)
    \biggr]
    +\ln (\tilde{\rho} ) \biggl[
            \frac{572 \tilde{\rho} }{45}
            -\frac{7133 \tilde{\rho} ^2}{90}
            \nonumber \\ &
            -\frac{56689 \tilde{\rho} ^3}{270}
            -\frac{8857 \tilde{\rho} ^4}{135}
            +\frac{56 \tilde{\rho} ^5}{75}
            +\frac{\tilde{\rho} ^6}{30}
            +\ln (1-\tilde{\rho} ) \biggl(
                    \frac{34}{45}-\frac{242 \tilde{\rho} }{9}-\frac{58 \tilde{\rho} ^2}{3}+\frac{1262 \tilde{\rho} ^3}{9}
                    \nonumber \\ &
                    +\frac{178 \tilde{\rho} ^4}{9}-\frac{14 \tilde{\rho} ^5}{5}-\frac{8}{3} \sqrt{\tilde{\rho} }-\frac{280}{9} \tilde{\rho} ^{3/2}+\frac{6104}{45} \tilde{\rho} ^{5/2}+120 \tilde{\rho} ^{7/2}\biggr)
    \biggr]
    \nonumber \\ &
    +\ln \left(\frac{\mu_s}{m_b}\right)
\biggl[-\frac{1}{2}
            -\frac{19 \tilde{\rho} }{3}
            +\frac{121 \tilde{\rho} ^2}{6}
            -\frac{52 \tilde{\rho} ^3}{3}
            +\frac{11 \tilde{\rho} ^4}{6}
            +\frac{7 \tilde{\rho} ^5}{3}
            -\frac{\tilde{\rho} ^6}{6}
            +\biggl(
                    -4 \tilde{\rho} +2 \tilde{\rho} ^2
                    \nonumber \\ &
                    +8 \tilde{\rho} ^3-6 \tilde{\rho} ^4\biggr) \ln (\tilde{\rho} )
    \biggr]
    +\ln \big(1+\sqrt{\tilde{\rho} }\big)
\biggl[-\frac{319}{2700}
            -\frac{1}{45 \tilde{\rho} }
            +\frac{6143 \tilde{\rho} }{540}
            +\frac{295 \tilde{\rho} ^2}{18}
            -\frac{365 \tilde{\rho} ^3}{27}
            \nonumber \\ &
            -\frac{6413 \tilde{\rho} ^4}{540}
            -\frac{659 \tilde{\rho} ^5}{300}
            -\frac{\tilde{\rho} ^6}{30}
            +\biggl(
                    -\frac{17}{45}+\frac{67 \tilde{\rho} }{9}+\frac{128 \tilde{\rho} ^2}{3}+\frac{404 \tilde{\rho} ^3}{9}+\frac{127 \tilde{\rho} ^4}{9}+\frac{7 \tilde{\rho} ^5}{5}\biggr) \ln (\tilde{\rho} )
    \biggr)
    \nonumber \\ &
    +\text{Li}_2\big(\sqrt{\tilde{\rho} }\big)
\biggl(\frac{32}{3} \sqrt{\tilde{\rho} }+\frac{1120}{9}
             \tilde{\rho} ^{3/2}-\frac{24416}{45} \tilde{\rho} ^{5/2}-480 \tilde{\rho} ^{7/2}\biggr)
    +\pi ^2 \biggl(
            -\frac{17}{180}+\frac{139 \tilde{\rho} }{36}-\frac{\tilde{\rho} ^2}{3}
            \nonumber \\ &
            -\frac{244 \tilde{\rho} ^3}{9}-\frac{161 \tilde{\rho} ^4}{36}+\frac{7 \tilde{\rho} ^5}{20}-\frac{4}{3} \sqrt{\tilde{\rho} }-\frac{140}{9} \tilde{\rho} ^{3/2}+\frac{3052}{45} \tilde{\rho} ^{5/2}+60 \tilde{\rho} ^{7/2}\biggr)
    +\text{Li}_2(\tilde{\rho} ) \biggl(
            \frac{17}{30}
            \nonumber \\ &
            -\frac{139 \tilde{\rho} }{6}+2 \tilde{\rho} ^2+\frac{488 \tilde{\rho} ^3}{3}+\frac{161 \tilde{\rho} ^4}{6}-\frac{21 \tilde{\rho} ^5}{10}-\frac{8}{3} \sqrt{\tilde{\rho} }-\frac{280}{9} \tilde{\rho} ^{3/2}+\frac{6104}{45} \tilde{\rho} ^{5/2}+120 \tilde{\rho} ^{7/2}\biggr)
            \nonumber \\ &
            +\biggl(
            5 \tilde{\rho} +\frac{\tilde{\rho} ^2}{6}-\frac{725 \tilde{\rho} ^3}{18}-\frac{305 \tilde{\rho} ^4}{36}+\frac{7 \tilde{\rho} ^5}{20}\biggr) \ln ^2(\tilde{\rho} )
\Biggr\}\,,
  \\
  Q_{3,0, \mu_G}^{(1)} &= C_A 
  \Biggl\{
    \frac{1147}{6480}
    +\frac{990487 \tilde{\rho} }{25200}
    -\frac{32728597 \tilde{\rho} ^2}{75600}
    -\frac{16174537 \tilde{\rho} ^3}{25200}
    +\frac{6247889 \tilde{\rho} ^4}{8400}
    +\frac{21711847 \tilde{\rho} ^5}{75600}
    \nonumber \\ &
    +\frac{963427 \tilde{\rho} ^6}{226800}
    +\frac{5 \tilde{\rho} ^7}{112}
    +\ln^2(\tilde{\rho} ) \biggl(
            \frac{29 \tilde{\rho} }{4}+\frac{289 \tilde{\rho} ^2}{12}-\frac{3925 \tilde{\rho} ^3}{36}-\frac{967 \tilde{\rho} ^4}{8}-\frac{34 \tilde{\rho} ^5}{3}+\frac{101 \tilde{\rho} ^6}{360}\biggr)
            \nonumber \\ &
    +\ln \big(1+\sqrt{\tilde{\rho} }\big)
\biggl[-\frac{263}{2520}
            -\frac{31}{2520 \tilde{\rho} }
            +\frac{4849 \tilde{\rho} }{300}
            +\frac{3997 \tilde{\rho} ^2}{72}
            -\frac{5 \tilde{\rho} ^3}{6}
            -\frac{5989 \tilde{\rho} ^4}{120}
            -\frac{3383 \tilde{\rho} ^5}{180}
            \nonumber \\ &
            -\frac{25283 \tilde{\rho} ^6}{12600}
            -\frac{\tilde{\rho} ^7}{56}
            +\ln (\tilde{\rho} ) \biggl(
                    -\frac{5}{18}+\frac{47 \tilde{\rho} }{5}+\frac{515 \tilde{\rho} ^2}{6}+\frac{1430 \tilde{\rho} ^3}{9}+\frac{179 \tilde{\rho} ^4}{2}+\frac{53 \tilde{\rho} ^5}{3}
                    \nonumber \\ &
                    +\frac{101 \tilde{\rho} ^6}{90}\biggr)
    \biggr]
    +\ln \big(1-\sqrt{\tilde{\rho} }\big)
\biggl[-\frac{263}{2520}
            -\frac{31}{2520 \tilde{\rho} }
            +\frac{4849 \tilde{\rho} }{300}
            +\frac{3997 \tilde{\rho} ^2}{72}
            -\frac{5 \tilde{\rho} ^3}{6}
            -\frac{5989 \tilde{\rho} ^4}{120}
            \nonumber \\ &
            -\frac{3383 \tilde{\rho} ^5}{180}
            -\frac{25283 \tilde{\rho} ^6}{12600}
            -\frac{\tilde{\rho} ^7}{56}
            +\ln (\tilde{\rho} ) \biggl(
                    -\frac{5}{18}+\frac{47 \tilde{\rho} }{5}+\frac{515 \tilde{\rho} ^2}{6}+\frac{1430 \tilde{\rho} ^3}{9}+\frac{179 \tilde{\rho} ^4}{2}
                    \nonumber \\ &
                    +\frac{53 \tilde{\rho} ^5}{3}+\frac{101 \tilde{\rho} ^6}{90}+\frac{16}{3} \sqrt{\tilde{\rho} }+\frac{1312}{9} \tilde{\rho} ^{3/2}-384 \tilde{\rho} ^{5/2}-\frac{7072}{5} \tilde{\rho} ^{7/2}-\frac{4016}{9} \tilde{\rho} ^{9/2}\biggr)
    \biggr]
    \nonumber \\ &
    +\ln (\tilde{\rho} ) \biggl[
            \frac{2473 \tilde{\rho} }{90}
            -\frac{6971 \tilde{\rho} ^2}{90}
            -\frac{762761 \tilde{\rho} ^3}{1080}
            -\frac{7781 \tilde{\rho} ^4}{12}
            -\frac{9007 \tilde{\rho} ^5}{90}
            +\frac{5659 \tilde{\rho} ^6}{6300}
            +\frac{\tilde{\rho} ^7}{56}
            \nonumber \\ &
            +\ln (1-\tilde{\rho} ) \biggl(
                    \frac{5}{9}-\frac{179 \tilde{\rho} }{5}-\frac{394 \tilde{\rho} ^2}{3}+\frac{3526 \tilde{\rho} ^3}{9}+434 \tilde{\rho} ^4+\frac{83 \tilde{\rho} ^5}{3}-\frac{101 \tilde{\rho} ^6}{45}-\frac{8}{3} \sqrt{\tilde{\rho} }
                    \nonumber \\ &
                    -\frac{656}{9} \tilde{\rho} ^{3/2}+192 \tilde{\rho} ^{5/2}+\frac{3536}{5} \tilde{\rho} ^{7/2}+\frac{2008}{9} \tilde{\rho} ^{9/2}\biggr)
    \biggr]
    +\ln \left(\frac{\mu_s}{m_b}\right)
\biggl[-\frac{23}{56}
            -\frac{573 \tilde{\rho} }{40}
            \nonumber \\ &
            +\frac{275 \tilde{\rho} ^2}{8}
            +\frac{315 \tilde{\rho} ^3}{8}
            -\frac{535 \tilde{\rho} ^4}{8}
            +\frac{49 \tilde{\rho} ^5}{8}
            +\frac{73 \tilde{\rho} ^6}{40}
            -\frac{5 \tilde{\rho} ^7}{56}
            +\biggl(
                    -6 \tilde{\rho} -\frac{15 \tilde{\rho} ^2}{2}+
                    \frac{105 \tilde{\rho} ^3}{2}
                    \nonumber \\ &
                    +\frac{45 \tilde{\rho} ^4}{2}-\frac{15 \tilde{\rho} ^5}{2}\biggr) \ln (\tilde{\rho} )
    \biggr]
    +\text{Li}_2\big(\sqrt{\tilde{\rho} }\big)
\biggl(\frac{32}{3} \sqrt{\tilde{\rho} }+\frac{2624}{9} \tilde{\rho} ^{3/2}-768 \tilde{\rho} ^{5/2}-\frac{14144}{5} \tilde{\rho} ^{7/2}
\nonumber \\ &
-\frac{8032}{9} \tilde{\rho} ^{9/2}\biggr)
    +\pi ^2 \biggl(
            -\frac{5}{72}+\frac{311 \tilde{\rho} }{60}+\frac{1061 \tilde{\rho} ^2}{72}-\frac{4241 \tilde{\rho} ^3}{54}-\frac{1915 \tilde{\rho} ^4}{24}-\frac{73 \tilde{\rho} ^5}{12}+\frac{101 \tilde{\rho} ^6}{360}
            \nonumber \\ &
            -\frac{4}{3} \sqrt{\tilde{\rho} }-\frac{328}{9} \tilde{\rho} ^{3/2}+96 \tilde{\rho} ^{5/2}+\frac{1768}{5} \tilde{\rho} ^{7/2}+\frac{1004}{9} \tilde{\rho} ^{9/2}\biggr)
    +\text{Li}_2(\tilde{\rho} ) \biggl(
            \frac{5}{12}-\frac{311 \tilde{\rho} }{10}
            \nonumber \\ &
            -\frac{1061 \tilde{\rho} ^2}{12}+\frac{4241 \tilde{\rho} ^3}{9}+\frac{1915 \tilde{\rho} ^4}{4}+\frac{73 \tilde{\rho} ^5}{2}-\frac{101 \tilde{\rho} ^6}{60}-\frac{8}{3} \sqrt{\tilde{\rho} }-\frac{656}{9} \tilde{\rho} ^{3/2}+192 \tilde{\rho} ^{5/2}
            \nonumber \\ &
            +\frac{3536}{5} \tilde{\rho} ^{7/2}+\frac{2008}{9} \tilde{\rho} ^{9/2}\biggr)
\Biggr\}
+C_F \Biggl\{
    -\frac{182927}{64800}
    -\frac{61448201 \tilde{\rho} }{352800}
    +\frac{1126861973 \tilde{\rho} ^2}{1058400}
    \nonumber \\ &
    +\frac{1267432219 \tilde{\rho} ^3}{1058400}
    -\frac{639220273 \tilde{\rho} ^4}{352800}
    -\frac{297234797 \tilde{\rho} ^5}{1058400}
    +\frac{26269529 \tilde{\rho} ^6}{3175200}
    -\frac{1567 \tilde{\rho} ^7}{2016}
    \nonumber \\ &
    +\ln ^2(\tilde{\rho} ) \biggl(
            -17 \tilde{\rho} -\frac{221 \tilde{\rho} ^2}{6}+\frac{3008 \tilde{\rho} ^3}{9}+\frac{2141 \tilde{\rho} ^4}{8}+\frac{53 \tilde{\rho} ^5}{24}+\frac{239 \tilde{\rho} ^6}{360}-\frac{5 \tilde{\rho} ^7}{56}\biggr)
    \nonumber \\ &
    +\ln \big(1+\sqrt{\tilde{\rho} }\big)
\biggl[\frac{8597}{5880}
            -\frac{4}{315 \tilde{\rho} }
            +\frac{3767 \tilde{\rho} }{600}
            -\frac{3907 \tilde{\rho} ^2}{72}
            +\frac{475 \tilde{\rho} ^3}{72}
            +\frac{9467 \tilde{\rho} ^4}{120}
            -\frac{11009 \tilde{\rho} ^5}{360}
            \nonumber \\ &
            -\frac{113107 \tilde{\rho} ^6}{12600}
            +\frac{237 \tilde{\rho} ^7}{392}
            +\ln (\tilde{\rho} ) \biggl(
                    \frac{67}{126}+\frac{111 \tilde{\rho} }{10}-\frac{155 \tilde{\rho} ^2}{6}-\frac{2135 \tilde{\rho} ^3}{18}+\frac{33 \tilde{\rho} ^4}{2}+\frac{317 \tilde{\rho} ^5}{6}
                    \nonumber \\ &
                    +\frac{239 \tilde{\rho} ^6}{90}-\frac{5 \tilde{\rho} ^7}{14}\biggr)
    \biggr]
    +\ln (\tilde{\rho} ) \biggl[
            -\frac{40357 \tilde{\rho} }{504}
            +\frac{473153 \tilde{\rho} ^2}{2520}
            +\frac{1549664 \tilde{\rho} ^3}{945}
            +\frac{974059 \tilde{\rho} ^4}{840}
            \nonumber \\ &
            +\frac{23911 \tilde{\rho} ^5}{210}
            +\frac{7823 \tilde{\rho} ^6}{900}
            -\frac{237 \tilde{\rho} ^7}{392}
            +\ln (1-\tilde{\rho} ) \biggl(
                    -\frac{67}{63}+\frac{369 \tilde{\rho} }{5}+\frac{340 \tilde{\rho} ^2}{3}-\frac{12404 \tilde{\rho} ^3}{9}
                    \nonumber \\ &
                    -1092 \tilde{\rho} ^4-
                    \frac{284 \tilde{\rho} ^5}{3}-\frac{239 \tilde{\rho} ^6}{45}+\frac{5 \tilde{\rho} ^7}{7}+\frac{32}{3} \sqrt{\tilde{\rho} }+\frac{1888}{9} \tilde{\rho} ^{3/2}-608 \tilde{\rho} ^{5/2}-\frac{59552}{35} \tilde{\rho} ^{7/2}
                    \nonumber \\ &
                    -\frac{3200}{9} \tilde{\rho} ^{9/2}\biggr)
    \biggr]
    +\ln \big(1-\sqrt{\tilde{\rho} }\big)
\biggl[\frac{8597}{5880}
            -\frac{4}{315 \tilde{\rho} }
            +\frac{3767 \tilde{\rho} }{600}
            -\frac{3907 \tilde{\rho} ^2}{72}
            +\frac{475 \tilde{\rho} ^3}{72}
            +\frac{9467 \tilde{\rho} ^4}{120}
            \nonumber \\ &
            -\frac{11009 \tilde{\rho} ^5}{360}
            -\frac{113107 \tilde{\rho} ^6}{12600}
            +\frac{237 \tilde{\rho} ^7}{392}
            +\ln (\tilde{\rho} ) \biggl(
                    \frac{67}{126}+\frac{111 \tilde{\rho} }{10}-\frac{155 \tilde{\rho} ^2}{6}-\frac{2135 \tilde{\rho} ^3}{18}+\frac{33 \tilde{\rho} ^4}{2}
                    \nonumber \\ &
                    +\frac{317 \tilde{\rho} ^5}{6}+\frac{239 \tilde{\rho} ^6}{90}-\frac{5 \tilde{\rho} ^7}{14}-\frac{64}{3} \sqrt{\tilde{\rho} }-\frac{3776}{9} \tilde{\rho} ^{3/2}+1216 \tilde{\rho} ^{5/2}+\frac{119104}{35} \tilde{\rho} ^{7/2}
                    \nonumber \\ &
                    +\frac{6400}{9} \tilde{\rho} ^{9/2}\biggr)
    \biggr]
    +\text{Li}_2(\tilde{\rho} ) \biggl(
            -\frac{67}{84}+\frac{1587 \tilde{\rho} }{20}+\frac{1205 \tilde{\rho} ^2}{12}-\frac{51751 \tilde{\rho} ^3}{36}-\frac{4335 \tilde{\rho} ^4}{4}-\frac{273 \tilde{\rho} ^5}{4}
            \nonumber \\ &
            -\frac{239 \tilde{\rho} ^6}{60}+\frac{15 \tilde{\rho} ^7}{28}+\frac{32}{3} \sqrt{\tilde{\rho} }+\frac{1888}{9} \tilde{\rho} ^{3/2}-608 \tilde{\rho} ^{5/2}-\frac{59552}{35} \tilde{\rho} ^{7/2}-\frac{3200}{9} \tilde{\rho} ^{9/2}\biggr)
    \nonumber \\ &
    +\pi ^2 \biggl(
            \frac{67}{504}-\frac{529 \tilde{\rho} }{40}-\frac{1205 \tilde{\rho} ^2}{72}+\frac{51751 \tilde{\rho} ^3}{216}+\frac{1445 \tilde{\rho} ^4}{8}+\frac{91 \tilde{\rho} ^5}{8}+\frac{239 \tilde{\rho} ^6}{360}-\frac{5 \tilde{\rho} ^7}{56}
            \nonumber \\ &
            +\frac{16}{3} \sqrt{\tilde{\rho} }+\frac{944}{9} \tilde{\rho} ^{3/2}-304 \tilde{\rho} ^{5/2}-\frac{29776}{35} \tilde{\rho} ^{7/2}-\frac{1600}{9} \tilde{\rho} ^{9/2}\biggr)
    +\text{Li}_2\big(\sqrt{\tilde{\rho} }\big)
\biggl(-\frac{128}{3} \sqrt{\tilde{\rho} }
\nonumber \\ &
-\frac{7552}{9} \tilde{\rho} ^{3/2}+2432 \tilde{\rho} ^{5/2}+\frac{238208}{35} \tilde{\rho} ^{7/2}+\frac{12800}{9} \tilde{\rho} ^{9/2}\biggr)
\Biggr\}\,,
  \\
  Q_{4,0, \mu_G}^{(1)} &= C_A \Biggl\{
    \frac{196879}{3175200}
    +\frac{24556379 \tilde{\rho} }{317520}
    -\frac{274033471 \tilde{\rho} ^2}{529200}
    -\frac{3087900503 \tilde{\rho} ^3}{1058400}
    +\frac{17682109 \tilde{\rho} ^4}{42336}
    \nonumber \\ &
    +\frac{1326250231 \tilde{\rho} ^5}{529200}
    +\frac{683258473 \tilde{\rho} ^6}{1587600}
    +\frac{11880109 \tilde{\rho} ^7}{3175200}
    +\frac{3 \tilde{\rho} ^8}{112}
    +\ln ^2(\tilde{\rho} ) \biggl(
            \frac{19 \tilde{\rho} }{2}+\frac{455 \tilde{\rho} ^2}{6}
            \nonumber \\ &
            -\frac{1468 \tilde{\rho} ^3}{9}-\frac{10741 \tilde{\rho} ^4}{18}-\frac{3977 \tilde{\rho} ^5}{15}-\frac{641 \tilde{\rho} ^6}{45}+\frac{74 \tilde{\rho} ^7}{315}\biggr)
    +\ln \big(1+\sqrt{\tilde{\rho} }\big)
\biggl[-\frac{2701}{29400}
\nonumber \\ &
            -\frac{19}{2520 \tilde{\rho} }
            +\frac{66539 \tilde{\rho} }{3150}
            +\frac{57841 \tilde{\rho} ^2}{450}
            +\frac{4942 \tilde{\rho} ^3}{45}
            -\frac{5201 \tilde{\rho} ^4}{45}
            -\frac{52157 \tilde{\rho} ^5}{450}
            -\frac{82039 \tilde{\rho} ^6}{3150}
            \nonumber \\ &
            -\frac{18079 \tilde{\rho} ^7}{9800}
            -\frac{3 \tilde{\rho} ^8}{280}
            +\ln (\tilde{\rho} ) \biggl(
                    -\frac{68}{315}+\frac{506 \tilde{\rho} }{45}+\frac{2246 \tilde{\rho} ^2}{15}+\frac{3974 \tilde{\rho} ^3}{9}+\frac{3862 \tilde{\rho} ^4}{9}
                    \nonumber \\ &
                    +\frac{2302 \tilde{\rho} ^5}{15}+\frac{946 \tilde{\rho} ^6}{45}+\frac{296 \tilde{\rho} ^7}{315}\biggr)
    \biggr]
    +\ln \big(1-\sqrt{\tilde{\rho} }\big)
\biggl[-\frac{2701}{29400}
            -\frac{19}{2520 \tilde{\rho} }
            +\frac{66539 \tilde{\rho} }{3150}
            \nonumber \\ &
            +\frac{57841 \tilde{\rho} ^2}{450}
            +\frac{4942 \tilde{\rho} ^3}{45}
            -\frac{5201 \tilde{\rho} ^4}{45}
            -\frac{52157 \tilde{\rho} ^5}{450}
            -\frac{82039 \tilde{\rho} ^6}{3150}
            -\frac{18079 \tilde{\rho} ^7}{9800}
            -\frac{3 \tilde{\rho} ^8}{280}
            \nonumber \\ &
            +\ln (\tilde{\rho} ) \biggl(
                    -\frac{68}{315}+\frac{506 \tilde{\rho} }{45}+\frac{2246 \tilde{\rho} ^2}{15}+\frac{3974 \tilde{\rho} ^3}{9}+\frac{3862 \tilde{\rho} ^4}{9}+\frac{2302 \tilde{\rho} ^5}{15}+\frac{946 \tilde{\rho} ^6}{45}
                    \nonumber \\ &
                    +\frac{296 \tilde{\rho} ^7}{315}+\frac{16}{3} \sqrt{\tilde{\rho} }+\frac{2320}{9} \tilde{\rho} ^{3/2}-\frac{8096}{45} \tilde{\rho} ^{5/2}-\frac{154528}{35} \tilde{\rho} ^{7/2}-\frac{194128}{45} \tilde{\rho} ^{9/2}
                    \nonumber \\ &
                    -\frac{6448}{9} \tilde{\rho} ^{11/2}\biggr)
    \biggr]
    +\ln (\tilde{\rho} ) \biggl[
            \frac{13949 \tilde{\rho} }{315}
            +\frac{524 \tilde{\rho} ^2}{21}
            -\frac{6215371 \tilde{\rho} ^3}{3780}
            -\frac{12309893 \tilde{\rho} ^4}{3780}
            \nonumber \\ &
            -\frac{4810361 \tilde{\rho} ^5}{3150}
            -\frac{31034 \tilde{\rho} ^6}{225}
            +\frac{27917 \tilde{\rho} ^7}{29400}
            +\frac{3 \tilde{\rho} ^8}{280}
            +\ln (1-\tilde{\rho} ) \biggl(
                    \frac{136}{315}-\frac{2002 \tilde{\rho} }{45}-\frac{5282 \tilde{\rho} ^2}{15}
                    \nonumber \\ &
                    +\frac{5266 \tilde{\rho} ^3}{9}+\frac{20558 \tilde{\rho} ^4}{9}+\frac{14506 \tilde{\rho} ^5}{15}+\frac{1618 \tilde{\rho} ^6}{45}-
                    \frac{592 \tilde{\rho} ^7}{315}-\frac{8}{3} \sqrt{\tilde{\rho} }-\frac{1160}{9} \tilde{\rho} ^{3/2}+\frac{4048}{45} \tilde{\rho} ^{5/2}
                    \nonumber \\ &
                    +\frac{77264}{35} \tilde{\rho} ^{7/2}+\frac{97064}{45} \tilde{\rho} ^{9/2}+\frac{3224}{9} \tilde{\rho} ^{11/2}\biggr)
    \biggr]
    +\ln \left(\frac{\mu_s}{m_b}\right)
\biggl[-\frac{97}{280}
            -\frac{2482 \tilde{\rho} }{105}
            +\frac{179 \tilde{\rho} ^2}{10}
            \nonumber \\ &
            +258 \tilde{\rho} ^3
            -\frac{290 \tilde{\rho} ^4}{3}
            -\frac{838 \tilde{\rho} ^5}{5}
            +\frac{109 \tilde{\rho} ^6}{10}
            +\frac{158 \tilde{\rho} ^7}{105}
            -\frac{3 \tilde{\rho} ^8}{56}
            +\biggl(
                    -8 \tilde{\rho} -33 \tilde{\rho} ^2+120 \tilde{\rho} ^3
                    \nonumber \\ &
                    +245 \tilde{\rho} ^4+48 \tilde{\rho} ^5-9 \tilde{\rho} ^6\biggr) \ln (\tilde{\rho} )
    \biggr]
    +\text{Li}_2\big(\sqrt{\tilde{\rho} }\big)
\biggl(\frac{32}{3} \sqrt{\tilde{\rho} }+\frac{4640}{9} \tilde{\rho} ^{3/2}-\frac{16192}{45} \tilde{\rho} ^{5/2}
\nonumber \\ &
-\frac{309056}{35} \tilde{\rho} ^{7/2}-\frac{388256}{45} \tilde{\rho} ^{9/2}-\frac{12896}{9} \tilde{\rho} ^{11/2}\biggr)
    +\pi ^2 \biggl(
            -\frac{17}{315}+\frac{583 \tilde{\rho} }{90}+\frac{4159 \tilde{\rho} ^2}{90}
            \nonumber \\ &
            -\frac{7253 \tilde{\rho} ^3}{54}-\frac{22489 \tilde{\rho} ^4}{54}-\frac{5219 \tilde{\rho} ^5}{30}-\frac{697 \tilde{\rho} ^6}{90}+\frac{74 \tilde{\rho} ^7}{315}-\frac{4}{3} \sqrt{\tilde{\rho} }-\frac{580}{9} \tilde{\rho} ^{3/2}+\frac{2024}{45} \tilde{\rho} ^{5/2}
            \nonumber \\ &
            +
            \frac{38632}{35} \tilde{\rho} ^{7/2}+\frac{48532}{45} \tilde{\rho} ^{9/2}+\frac{1612}{9} \tilde{\rho} ^{11/2}\biggr)
    +\text{Li}_2(\tilde{\rho} ) \biggl(
            \frac{34}{105}-\frac{583 \tilde{\rho} }{15}-\frac{4159 \tilde{\rho} ^2}{15}
            \nonumber \\ &
            +\frac{7253 \tilde{\rho} ^3}{9}+\frac{22489 \tilde{\rho} ^4}{9}+\frac{5219 \tilde{\rho} ^5}{5}+\frac{697 \tilde{\rho} ^6}{15}-\frac{148 \tilde{\rho} ^7}{105}-\frac{8}{3} \sqrt{\tilde{\rho} }-\frac{1160}{9} \tilde{\rho} ^{3/2}+\frac{4048}{45} \tilde{\rho} ^{5/2}
            \nonumber \\ &
            +\frac{77264}{35} \tilde{\rho} ^{7/2}+\frac{97064}{45} \tilde{\rho} ^{9/2}+\frac{3224}{9} \tilde{\rho} ^{11/2}\biggr)
\Biggr\}
+C_F \Biggl\{
    -\frac{6339713}{2540160}
    -\frac{26167927 \tilde{\rho} }{75600}
    \nonumber \\ &
    +\frac{1186820717 \tilde{\rho} ^2}{1058400}
    +\frac{1265814329 \tilde{\rho} ^3}{176400}
    -\frac{584924819 \tilde{\rho} ^4}{264600}
    -\frac{2839397629 \tilde{\rho} ^5}{529200}
    \nonumber \\ &
    -\frac{1204660967 \tilde{\rho} ^6}{3175200}
    +\frac{783799 \tilde{\rho} ^7}{105840}
    -\frac{44797 \tilde{\rho} ^8}{94080}
    +\ln ^2(\tilde{\rho} ) \biggl(
            -22 \tilde{\rho} -\frac{445 \tilde{\rho} ^2}{3}+\frac{5962 \tilde{\rho} ^3}{9}
            \nonumber \\ &
            +\frac{3439 \tilde{\rho} ^4}{2}+\frac{8633 \tilde{\rho} ^5}{15}+\frac{19 \tilde{\rho} ^6}{9}+\frac{59 \tilde{\rho} ^7}{105}-\frac{3 \tilde{\rho} ^8}{56}\biggr)
    +\ln \big(1+\sqrt{\tilde{\rho} }\big)
\biggl[\frac{240929}{176400}
            -\frac{1}{126 \tilde{\rho} }
            \nonumber \\ &
            +\frac{36973 \tilde{\rho} }{2205}
            -\frac{15911 \tilde{\rho} ^2}{225}
            -\frac{917 \tilde{\rho} ^3}{5}
            +\frac{1589 \tilde{\rho} ^4}{9}
            +\frac{26719 \tilde{\rho} ^5}{225}
            -\frac{16339 \tilde{\rho} ^6}{315}
            -\frac{172847 \tilde{\rho} ^7}{22050}
            \nonumber \\ &
            +\frac{4609 \tilde{\rho} ^8}{11760}
            +\ln (\tilde{\rho} ) \biggl(
                    \frac{289}{630}+\frac{356 \tilde{\rho} }{21}+\frac{28 \tilde{\rho} ^2}{15}-\frac{2828 \tilde{\rho} ^3}{9}-350 \tilde{\rho} ^4+\frac{812 \tilde{\rho} ^5}{15}+\frac{580 \tilde{\rho} ^6}{9}
                    \nonumber \\ &
                    +\frac{236 \tilde{\rho} ^7}{105}-\frac{3 \tilde{\rho} ^8}{14}\biggr)
    \biggr]
    +\ln (\tilde{\rho} ) \biggl[
            -\frac{83899 \tilde{\rho} }{630}
            -\frac{79687 \tilde{\rho} ^2}{630}
            +\frac{7931981 \tilde{\rho} ^3}{1890}
            +\frac{1334683 \tilde{\rho} ^4}{180}
            \nonumber \\ &
            +\frac{8494769 \tilde{\rho} ^5}{3150}
            +\frac{19645 \tilde{\rho} ^6}{126}
            +\frac{83011 \tilde{\rho} ^7}{11025}
            -\frac{4609 \tilde{\rho} ^8}{11760}
            +\ln (1-\tilde{\rho} ) \biggl(
                    -\frac{289}{315}+\frac{1976 \tilde{\rho} }{21}
                    \nonumber \\ &
                    +\frac{8354 \tilde{\rho} ^2}{15}-\frac{25816 \tilde{\rho} ^3}{9}-6679 \tilde{\rho} ^4-\frac{34384 \tilde{\rho} ^5}{15}-\frac{1034 \tilde{\rho} ^6}{9}-\frac{472 \tilde{\rho} ^7}{105}+\frac{3 \tilde{\rho} ^8}{7}+\frac{32}{3} \sqrt{\tilde{\rho} }
                    \nonumber \\ &
                    +\frac{3584}{9} \tilde{\rho} ^{3/2}-\frac{6464}{15} \tilde{\rho} ^{5/2}-\frac{43648}{7} \tilde{\rho} ^{7/2}
                    -
                    \frac{1581536}{315} \tilde{\rho} ^{9/2}-\frac{1664}{3} \tilde{\rho} ^{11/2}\biggr)
    \biggr]
    \nonumber \\ &
    +\ln \big(1-\sqrt{\tilde{\rho} }\big)
\biggl[\frac{240929}{176400}
            -\frac{1}{126 \tilde{\rho} }
            +\frac{36973 \tilde{\rho} }{2205}
            -\frac{15911 \tilde{\rho} ^2}{225}
            -\frac{917 \tilde{\rho} ^3}{5}
            +\frac{1589 \tilde{\rho} ^4}{9}
            \nonumber \\ &
            +\frac{26719 \tilde{\rho} ^5}{225}
            -\frac{16339 \tilde{\rho} ^6}{315}
            -\frac{172847 \tilde{\rho} ^7}{22050}
            +\frac{4609 \tilde{\rho} ^8}{11760}
            +\ln (\tilde{\rho} ) \biggl(
                    \frac{289}{630}+\frac{356 \tilde{\rho} }{21}+\frac{28 \tilde{\rho} ^2}{15}
                    \nonumber \\ &
                    -\frac{2828 \tilde{\rho} ^3}{9}-350 \tilde{\rho} ^4+\frac{812 \tilde{\rho} ^5}{15}+\frac{580 \tilde{\rho} ^6}{9}+\frac{236 \tilde{\rho} ^7}{105}-\frac{3 \tilde{\rho} ^8}{14}-\frac{64}{3} \sqrt{\tilde{\rho} }-\frac{7168}{9} \tilde{\rho} ^{3/2}
                    \nonumber \\ &
                    +\frac{12928}{15} \tilde{\rho} ^{5/2}+\frac{87296}{7} \tilde{\rho} ^{7/2}+\frac{3163072}{315} \tilde{\rho} ^{9/2}+\frac{3328}{3} \tilde{\rho} ^{11/2}\biggr)
    \biggr]
    +\text{Li}_2(\tilde{\rho} ) \biggl(
            -\frac{289}{420}
            \nonumber \\ &
            +\frac{718 \tilde{\rho} }{7}+\frac{8368 \tilde{\rho} ^2}{15}-\frac{27230 \tilde{\rho} ^3}{9}-6854 \tilde{\rho} ^4-\frac{11326 \tilde{\rho} ^5}{5}-\frac{248 \tilde{\rho} ^6}{3}-\frac{118 \tilde{\rho} ^7}{35}+\frac{9 \tilde{\rho} ^8}{28}
            \nonumber \\ &
            +\frac{32}{3} \sqrt{\tilde{\rho} }+\frac{3584}{9} \tilde{\rho} ^{3/2}
            -\frac{6464}{15} \tilde{\rho} ^{5/2}-\frac{43648}{7} \tilde{\rho} ^{7/2}-\frac{1581536}{315} \tilde{\rho} ^{9/2}-\frac{1664}{3} \tilde{\rho} ^{11/2}\biggr)
            \nonumber \\ &
    +\pi ^2 \biggl(
            \frac{289}{2520}-\frac{359 \tilde{\rho} }{21}-\frac{4184 \tilde{\rho} ^2}{45}+\frac{13615 \tilde{\rho} ^3}{27}+\frac{3427 \tilde{\rho} ^4}{3}+\frac{5663 \tilde{\rho} ^5}{15}+\frac{124 \tilde{\rho} ^6}{9}+\frac{59 \tilde{\rho} ^7}{105}
            \nonumber \\ &
            -\frac{3 \tilde{\rho} ^8}{56}+\frac{16}{3} \sqrt{\tilde{\rho} }+\frac{1792}{9} \tilde{\rho} ^{3/2}-\frac{3232}{15} \tilde{\rho} ^{5/2}-\frac{21824}{7} \tilde{\rho} ^{7/2}-\frac{790768}{315} \tilde{\rho} ^{9/2}-\frac{832}{3} \tilde{\rho} ^{11/2}\biggr)
            \nonumber \\ &
            +\text{Li}_2\big(\sqrt{\tilde{\rho} }\big)
\biggl(-\frac{128}{3} \sqrt{\tilde{\rho} }-\frac{14336}{9} \tilde{\rho} ^{3/2}+\frac{25856}{15} \tilde{\rho} ^{5/2}+\frac{174592}{7} \tilde{\rho} ^{7/2}+\frac{6326144}{315} \tilde{\rho} ^{9/2}
\nonumber \\ &
+\frac{6656}{3} \tilde{\rho} ^{11/2}\biggr)
\Biggr\}\,,
\end{align}
\begin{align}
  Q_{i,0, \tilde{\rho}_{LS}}^{(1)} &= - Q_{i,0, \mu_G}^{(1)} - \frac{1}{2} \left(C_A \ln \left(\frac{\mu _s}{m_b}\right)+C_A+C_F\right) Q_{i,0, \mu_G}^{(0)}\,,
\end{align}
where $\tilde{\rho}=m_c^2/m_b^2$.
The results presented in this Appendix are obtained 
from the differential expressions of Ref.~\cite{Mannel:2021zzr}
after integration over the dilepton pair invariant mass squared.

The analytic expressions shown in the Appendix can also
be obtained from~\cite{progdata}.



\bibliographystyle{JHEP} 
\footnotesize
\bibliography{BIB}

\end{document}